\documentclass{roajarticle}
\usepackage{graphicx}
\usepackage{mynatbib}
\usepackage{upgreek}
\usepackage{upgreek2}
\usepackage{amsmath}
\usepackage[section]{placeins}
\newcommand{\volume}{\textbf{1}}
\newcommand{\numarul}{1}
\newcommand{\volyear}{2019}
\graphicspath{ {./figs/} }

\def\be{\begin{equation}}
\def\ee{\end{equation}}
\def\bea{\begin{eqnarray}}
\def\eea{\end{eqnarray}}

\title{A Brief Introduction to the Adomian Decomposition Method, with Applications in Astronomy and Astrophysics}
\newcommand{\shorttitle}{The Adomian Decomposition Method with Applications}

\author[1]{Man Kwong \MakeTextUppercase{Mak}}
\author[2]{Chun Sing \MakeTextUppercase{Leung}}
\author[3] {Tiberiu \MakeTextUppercase{Harko}}
\affil[1]{Departamento de F\'{\i}sica, Facultad de Ciencias Naturales, Universidad de
Atacama, Copayapu 485, Copiap\'o, Chile,
Email: mankwongmak.@gmail.com}

\affil[2] {Department of Mathematics, Polytechnic University of Hong Kong, Hong Kong, Email: chun-sing-hkpu.leung@polyu.edu.hk}

\affil[3]{School of Physics, Sun Yat-Sen University, Xingang Road, Guangzhou 510275, People's Republic of China, Email: tiberiu.harko@aira.astro.ro}

\keywords{Mathematical Methods in Physics -- Ordinary Nonlinear Differential Equations -- Celestial Mechanics -- Astronomy--General Relativity}

\begin{document}
\maketitle

\begin{abstract}
The Adomian Decomposition Method (ADM) is a very effective approach for solving broad classes of nonlinear  partial and ordinary differential equations, with important applications in different fields of applied mathematics, engineering, physics and biology. It is the goal of the present paper to provide a clear and pedagogical introduction to the Adomian Decomposition Method and to some of its applications. In particular,  we focus our attention to a number of standard first-order ordinary differential equations (the linear, Bernoulli, Riccati, and Abel) with arbitrary coefficients, and present in detail the Adomian method for obtaining their solutions. In each case we compare the Adomian solution with the exact solution of some particular differential equations, and we show their complete equivalence. The second order and the fifth order ordinary differential equations are also considered. An important extension of the standard ADM, the Laplace-Adomian Decomposition Method is also introduced through the investigation of the solutions of a specific second order nonlinear differential equation. We also present the applications of the method to the Fisher-Kolmogorov second order partial nonlinear differential equation, which plays an important role in the description of many physical processes, as well as three important applications in astronomy and astrophysics, related to the determination of the solutions of the Kepler equation, of the Lane-Emden equation, and of the general relativistic equation describing the motion of massive particles in the spherically symmetric and static Schwarzschild geometry. \\
\end{abstract}

\tableofcontents

\section{Introduction}

In recent years, a lot of consideration has been dedicated to the investigations of the
Adomian's Decomposition Method (ADM) \citep{new0, R1,R2,b3, b2, b4}, which allows us to explore the solutions and properties of a large variety of ordinary and partial differential equations, as well as of integral equations, which describe various mathematical problems, or can be used to mathematically model diverse physical processes.  From a historical point of view, the ADM was first
introduced, and extensively used in the 1980's \citep{new0, new2,new3,new4, new1}, and ever since many mathematicians and scientists have continuously modified the ADM
in an attempt to enhance its accuracy and/or to broaden the applications of
the initial method \citep{b3, b2,b5,b6,b7a,b8,b9,b10,b11,b12,b13,b14,b28,b30,p1,p2,p3, b32, b31}.

An important benefit of the Adomian Decomposition Method is that it can yield
analytical approximations to quite extensive classes of nonlinear (and
stochastic) differential equations without resorting to discretization,  perturbation, linearization, or closure approximations
 methods, which could result in the necessity of extensive numerical
computations. For most of the mathematical models used for the mathematical description of natural phenomena, in order to obtain the analytical solutions of a nonlinear problem in a closed-form, and thus to make it solvable, it is usually necessary to make some simplifying assumptions, or t impose some restrictive conditions.

It is worth to note that ADM can provide a solution of a differential/integral equation
in the form of a series, whose terms are determined individually step by step via a recursive relation
using the Adomian polynomials. The main advantage of the Adomian Decomposition Method is
that the series solution of the differential/integral equation converges very quickly \citep{C1,C2, b3}, and
therefore it saves significant amounts of computing time. On the other hand, it is important to point out again that in the Adomian Decomposition Method there is no need
to  discretize or linearize the differential and integral equations. One
can find reviews of ADM in applied mathematics, and its applications in
science in \citet{R1}, \citet{R2}, and \citet{Haldar}, respectively.

The basic nonlinear ordinary differential equations of mathematics (Riccati and Abel), as well as their physical and engineering applications have continuously attracted the interest of mathematicians and physicists \citep{Mak1,Mak2,Mak3,Mak4,Mak5,Mak6,Mak7,Mak8}. These equations also proved to be a fertile investigation ground from the point of view of the ADM approach.  Recently, using the ADM, the Riccati equation was solved in \citet
{Ricc}. The Abel differential equation, having constant coefficients, of the
form
\be
\frac{dy}{dt}=\sum_{k=0}^{M}f_{k}y^{k},
\ee
 was solved with the help of
ADM in \citet{AB}. A modified version of the Adomian Decomposition Mthod was introduced for solving second order
ordinary differential equation in \citet{YL} and \citet{b15}, respectively. A
particular third order ordinary differential equation  was investigated by using a modified ADM for solving it in
\citet{PP}. The ADM was applied to the third order ordinary differential
equation
\be
y^{\prime \prime \prime }=y^{-k},
\ee
 representing a particular case of a generalized thin film equation describing the flow of a thin film downward of a vertical wall \citet{EM}. The ADM
for solving different classes of differential equations of importance in mathematical physics was studied in
\citet{PN}. The fourth order differential equation was solved by ADM in \citet%
{BB}. The biharmonic nonlinear Schr\"{o}dinger equation, and its standing wave solutions were investigated, via the use of the Laplace-Adomian and Adomian Decomposition Methods, in \citet{HMB}.

The Adomian Decomposition Method method was extensively applied in different areas of science and technology, including the study of the dynamics of the population growth models, which can be modelled  by single partial or ordinary differential equations, or
complex systems of such equations.  A few example of specific mathematical systems successfully explored by using the ADM are the shallow water waves \citep{swave}, the Brusselator model \citep{Brus}, the Lotka-Volterra model \citep{LV}, and the Belousov-Zhabotinsky reduction model \citep{BJ}, respectively.  The Adomian Decomposition Method was applied for the study of the Susceptible-Infected-Recovered (SIR) epidemic model, which is widely applied for the study of the spread of infectious diseases, in \citet{S1} and \citet{S2}, respectively.

 The Adomian Decomposition Method did also find some important applications in Physics. Nonlinear matrix differential equations of a new type, which emerge in general relativity as well as other scientific fields, were investigated in \citet{Az}. The solution of the nonlinear Klein-Gordon equation was obtained via the Adomian Decomposition Method in \citet{Gha}. The obtained semi-analytical solutions are in good accord with the full numerical solutions. The equations of motion of the massive and massless particles in the spherically symmetric and static Schwarzschild geometry of general relativity  were studied extensively in \citet{PP} by using the Laplace-Adomian Decomposition Method.  The physical properties of vortices with arbitrary topological charges arising in weakly interacting Bose-Einstein Condensates, described by differential equations of the form
 \be
 \frac{d^2R(x)}{dx^2}+\frac{1}{x}\frac{dR(x)}{dx}-\left[\frac{l^2}{x^2}+\left(v(x)-1\right)\right]R(x)-R^3(x)=0,
 \ee
  where $l$ is a constant, and $v(x)=0$ and $v(x)=x^2$,   were investigated using the Adomian Decomposition Method  in \citet{V}, where the nonlinear Gross-Pitaevskii equation was solved in polar coordinates. Series solutions using the Adomian Decomposition Method have been obtained for the Schr\"{o}dinger-Newton-$\Lambda$ system, described by the system of partial differential equations,
\begin{equation}
i\hbar \frac{\partial \psi \left( \vec{r},t\right) }{\partial t}=-\frac{%
\hbar ^{2}}{2m}\nabla^2 \psi \left( \vec{r},t\right) +m\Phi \left( \vec{r}%
,t\right) \psi \left( \vec{r},t\right) ,
\end{equation}%
\begin{equation}
\nabla^2 \Phi \left( \vec{r},t\right) =4\pi Gm\left\vert \psi \left( \vec{r}%
,t\right) \right\vert ^{2}-\frac{1}{2}\Lambda c^{2},
\end{equation}
where by $\psi \left( \vec{r},t\right)$ we have denoted  the particle wave function, $\Phi \left( \vec{r},t\right)$ is the gravitational potential, $\hbar$, $G$ and $\Lambda$ are the Planck, the gravitational and the cosmological constants, respectively,  while $m$ is the mass of the particle, in \citet{SN1} and \citet{SN2}, respectively.

Despite the existence of a large literature on the ADM, to the best knowledge of the authors no clearly written and pedagogical introduction to the method, which would be useful for a large audience of scientists from different fields, does exist presently.
It is the purpose of the present paper to give such  an introductory review of the Adomian Decomposition Method, and of the Laplace-Adomian Decomposition Method, in which, by means of the detailed and explicit presentation of all the calculations, and by providing a large number of examples, the power and efficiency of the method is clearly outlined. Hopefully, such a presentation would be of interest even for undergraduate students studying sciences and engineering, and will determine them to proceed to the study and investigation of the advanced features of the method.

From the point of view of the applications of the Adomian Decomposition Method in science we have chosen to present the analysis of the Fisher-Kolmogorov equation \citep{Fish,Kolm}, which plays an essential role in many physical and biological problems. But the main focus of the present paper are the potential astronomical and astrophysical applications of the Adomian Decomposition Method, a field that has yet to be explored in detail. One important astronomical problem that can be handled efficiently and effectively with the Adomian Decomposition Method is obtaining the solution of the Kepler equation, which plays a fundamental role in the determination of the orbits of the celestial orbits. The hyperbolic and the elliptic Kepler equations were investigated by using ADM in \citet{Ebaid} and \citet{Ebaid1}, respectively. One of the basic equations of Newtonian astrophysics is the Lane-Emden equation, which was used, for example, for the study of the white dwarfs, which lead to the fundamental Chandrasekhar mass limit for this type of compact objects  \citep{Chan}. The Lane-Emden equation was intensively investigated by using the Adomian Decomposition Method, which provides an efficient and computationally powerful procedure to obtain its solutions, in \citep{Ad95, LE0, LE0a, LEa, LE1}. Finally, we will consider the general relativistic motion of massive test particles in the static and spherically symmetric Schwarzschild geometry, and present its Adomian series solution \citep{PP}. This approach can be  used for the extremely precise analytical calculation of the orbit of the planet Mercury, for the study of its perihelion precession, as well as for the computation of the light deflection by the Sun. The solutions of the Kompaneets equation, a nonlinear partial differential equation that plays
an important role in astrophysics, describing the spectra of photons in interaction with a
rarefied electron gas, were obtained, by using the Laplace-Adomian Decomposition Method, in \citet{Gax}.

The present paper is organized as follows. We introduce the basics of the Adomian Decomposition Method in Section~\ref{sect20}. In Section~\ref{sect2} we discuss the application of the ADM to the case of the first order differential equations. We begin our discussion with the simplest case of the ordinary linear first order differential equation, whose solution can be obtained exactly. The power series solution of the linear equation is obtained by using a power series expansion. We consider then a particular case, and we compare the power series and the exact solutions. Next we proceed to the investigation of the Bernoulli, Riccati and Abel type equations with constant coefficients, by using ADM, and the series solutions of these equations are obtained. In each case the power series solution is compared with the exact solution of a particular differential equation. The case of the second order differential equations is considered in Section~\ref{sect3}. Two specific example are also presented, and discussed in detail. The fifth order ordinary differential equation is analyzed in Section~\ref{sect4}. As an example of the use of the ADM for solving nonlinear partial differential equations, in Section~\ref{sect5} we consider the case of the Fisher-Kolmogorov equation, a nonlinear differential equation with many applications in biology. We present the Laplace-Adomian Decomposition Method for second order nonlinear differential equations in Section~\ref{sect1n}. Astronomical and astrophysical applications of the Adomian Decomposition Method (Kepler equation, Lane-Emden equation, and the motion of massive particles in the Schwarzschild geometry) are presented in Section~\ref{sect2n}.  Finally, we discuss and conclude our results in Section~\ref{sect6}.

\section{The Adomian Decomposition Method}\label{sect20}

We illustrate now the basic ideas of the Adomian Decomposition Method by considering the case of a nonlinear partial differential equation written in the general form
\begin{equation}
\hat{L}_{t}\left[ y\left( x,t\right) \right] +\hat{R}\left[ y\left(
x,t\right) \right] +\hat{N}\left[ y\left( x,t\right) \right] =f(x,t),
\label{0}
\end{equation}%
where $\hat{L}_{t}\left[ .\right] =\partial /\partial t\left[ .\right] $ denotes the partial derivative operator with respect to the time $t$, while $%
\hat{R}\left[ .\right] $ is the linear operator, generally containing
partial derivatives with respect to $x$. Moreover,  $\hat{N}\left[ .\right] $ represents a
nonlinear analytic operator, and $f(x,t)$ is a non-homogeneous arbitrary function,
assumed to be independent of $y(x,t)$. Eq.~(\ref{0}) has to be considered together with the
initial condition $y(x,0)=g(x)$. In the following w assume that the operator $\hat{L}_{t}$ is invertible,
and therefore we can apply $\hat{L}_{t}^{-1}$ to both sides of Eq.~(\ref{0}), thus first obtaining
\begin{eqnarray}
y(x,t)&=&g(x)+\hat{L}_{t}^{-1}\left[ f(x,t)\right]-
\hat{L}_{t}^{-1}\hat{R}%
\left[ y\left( x,t\right) \right] - \hat{L}_{t}^{-1}\hat{N}\left[ y\left( x,t\right) \right] .
\end{eqnarray}

The ADM postulates the existence of a series solution of Eq.~(\ref{0}) in which $y(x,t)$ can be represented by
\begin{equation}
y(x,t)=\sum_{n=0}^{\infty }y_{n}\left( x,t\right) .  \label{01}
\end{equation}%
Moreover, it is assumed  that the nonlinear term $\hat{N}\left[ y\left( x,t\right) \right] $ can be
decomposed according to
\begin{equation}
\hat{N}\left[ y\left( x,t\right) \right] =\sum_{n=0}^{\infty }A_{n}\left(
y_{0},y_{1},...,y_{n}\right) ,  \label{02}
\end{equation}%
where $\left\{ A_{n}\right\} _{n=0}^{\infty }$ are called the Adomian polynomials.
They can be computed according to the simple rule  \citep{new0, R1,R2,b3, b2}
\begin{equation}
A_{n}\left( y_{0},y_{1},...,y_{n}\right) =\frac{1}{n!}\frac{d^{n}}{d\epsilon
^{n}}\hat{N}\left(t,\sum_{k=0}^{n}\epsilon^{k}y_{k}\right)
\Bigg|_{\epsilon =0}.
\end{equation}

After the substitution of the series expansions (\ref{01}) and (\ref{02}) into Eq. (\ref%
{0}), we obtain
\begin{eqnarray}
\sum_{n=0}^{\infty }y_{n}\left( x,t\right) &=&g(x)+\hat{L}_{t}^{-1}\left[
f(x,t)\right] \hat{L}_{t}^{-1}\hat{R}\left[ \sum_{n=0}^{\infty }y_{n}\left( x,t\right) %
\right] \notag \\
&-&\hat{L}_{t}^{-1}\left[ \sum_{n=0}^{\infty }A_{n}\left(
y_{0},y_{1},...,y_{n}\right) \right] .
\end{eqnarray}

From the above equation we immediately obtain the following recurrence relation, which gives the series
solution of Eq. (\ref{0}) as
\begin{equation}
y_{0}\left( x,t\right) =g(x)+\hat{L}_{t}^{-1}\left[ f(x,t)\right] ,
\end{equation}
\begin{eqnarray}
y_{k+1}(x,t)&=&\hat{L}_{t}^{-1}\hat{R}\left[ y_{k}\left( x,t\right) \right] -%
\hat{L}_{t}^{-1}\left[ A_{k}\left( y_{0},y_{1},...,y_{n}\right) \right],
\notag\\
&&k=0,1,2, \dots
\end{eqnarray}

Therefore, an approximate solution of Eq.~(\ref{0}) is obtained as
\begin{equation}
y(x,t)\simeq \sum_{k=0}^{n}y_{k}\left( x,t\right) ,
\end{equation}
and
\begin{equation}
\lim_{n\rightarrow\infty }\sum_{k=0}^{n}y_{k}\left( x,t\right) =y(x,t).
\end{equation}

For an arbitrary nonlinearity $\hat{N}\left[ y(x,t)%
\right] $, the Adomian polynomials can be obtained according to the rule
\begin{equation}
A_{0}=\hat{N}\left[ y_{0}\right] , \quad A_{1}=y_{1}\frac{d}{dy_{0}}\hat{N}\left[
y_{0}\right] ,
\end{equation}
\begin{equation}
A_{2}=y_{2}\frac{d}{dy_{0}}\hat{N}\left[ y_{0}\right] +\frac{y_{1}^{2}}{2!}%
\frac{d^{2}}{dy_{0}^{2}}\hat{N}\left[ y_{0}\right] ,
\end{equation}
\begin{equation}
A_{3}=y_{3}\frac{d}{dy_{0}}\hat{N}\left[ y_{0}\right] +y_{1}y_{2}\frac{d^{2}%
}{dy_{0}^{2}}\hat{N}\left[ y_{0}\right] +\frac{y_{1}^{3}}{3!}\frac{d^{3}}{%
dy_{0}^{3}}\hat{N}\left[ y_{0}\right] .
\end{equation}
This procedure can be continued indefinitely.  The greater the number of considered  terms in the Adomian Decomposition Method series expansion, the higher is the numerical accuracy of the semi-analytical solution.

In the following Sections we will present in detail the application of the Adomian Decomposition Method for a large class of nonlinear ordinary and partial differential equations.

\section{The Adomian
Decomposition Method for first order ordinary differential equations}\label{sect2}

In the present Section we introduce the application of the ADM to the case of first order differential equations. The linear, Bernoulli, Riccati and Abel differential equations are considered in detail.

\subsection{Linear differential equation $\frac{dy}{dx}+P\left( x\right)
y=Q\left( x\right) $}

The decomposition method can be used to solve the linear differential
equations. Consider that the differential equation takes the standard form of the first order ordinary differential equation,

\begin{equation}
\frac{dy}{dx}+P\left( x\right) y=Q\left( x\right) ,  \label{1}
\end{equation}
where $P\left( x\right) $ and $Q\left( x\right) $ are arbitrary function of $%
x$. Eq.~(\ref{1}) must be solved together with the initial condition $y\left( 0\right)=y_0 $. Assume that the solution of Eq. (\ref{1}) can be obtained in power series
form,
\begin{equation}
y\left( x\right) =\sum_{n=0}^{\infty }y_{n}\left( x\right) .  \label{2}
\end{equation}

Now integrating Eq. (\ref{1}) yields the integral equation%
\begin{equation}
y\left( x\right) =y\left( 0\right) +\int_{0}^{x}Q\left( x\right)
dx-\int_{0}^{x}P\left( x\right) ydx.  \label{3}
\end{equation}

 Substituting Eq. (\ref{2}%
) into Eq. (\ref{3}) gives the relation%
\begin{eqnarray}
\sum_{n=0}^{\infty }y_{n}\left( x\right) &=&y_{0}\left( x\right)
+\sum_{n=1}^{\infty }y_{n}\left( x\right) =y_{0}\left( x\right)
+\sum_{n=0}^{\infty }y_{n+1}\left( x\right)  \notag \\
&=&y\left( 0\right) +\int_{0}^{x}Q\left( x\right) dx-\int_{0}^{x}P\left(
x\right) \sum_{n=0}^{\infty }y_{n}\left( x\right) dx.  \label{4}
\end{eqnarray}

Next we rewrite Eq. (\ref{4}) in the recursive forms%
\begin{equation}
y_{0}\left( x\right) =y\left( 0\right) +\int_{0}^{x}Q\left( x\right) dx,
\label{5}
\end{equation}%
\begin{equation}
y_{k+1}\left( x\right) =-\int_{0}^{x}P\left( x\right) y_{k}\left( x\right)
dx.  \label{6}
\end{equation}%

From Eqs. (\ref{5}) and (\ref{6}) we can obtain
the approximate semi-analytical solution of  Eq. (\ref{1}), as given by
\be
y\left( x\right)
=\sum_{n=0}^{\infty }y_{n}\left( x\right).
\ee

\subsubsection{Example: $\frac{dy}{dx}+2xy=4x^{3}$}

Consider the differential equation %
\begin{equation}
\frac{dy}{dx}+2xy=4x^{3},
\end{equation}
which we solve with the initial condition $y\left( 0\right) =1$.
Then its general solution is given by%
\begin{equation}\label{ex1}
y\left( x\right) =3e^{-x^{2}}+2\left( x^{2}-1\right).
\end{equation}
In the present case we have  $P\left(
x\right) =2x$ and $Q\left( x\right) =4x^{3}$, respectively. Hence the power series of the equation is obtained as
\begin{equation}
y_{0}\left( x\right) =y\left( 0\right) +\int_{0}^{x}Q\left( x\right)
dx=1+x^{4},
\end{equation}%
\begin{equation}
y_{1}\left( x\right) =-2\int_{0}^{x}xy_{0}\left( x\right) dx=-x^{2}-\frac{%
x^{6}}{3},
\end{equation}%
\begin{equation}
y_{2}\left( x\right) =-2\int_{0}^{x}xy_{1}\left( x\right) dx=\frac{x^{4}}{2}+%
\frac{x^{8}}{12},
\end{equation}%
\begin{equation}
y_{3}\left( x\right) =-2\int_{0}^{x}xy_{2}\left( x\right) dx=-\frac{x^{6}}{3}%
-\frac{x^{10}}{60},
\end{equation}%
\begin{equation}
y_{4}\left( x\right) =-2\int_{0}^{x}xy_{3}\left( x\right) dx=\frac{x^{8}}{24}%
+\frac{x^{12}}{360}.
\end{equation}

\begin{equation}
y\left( x\right) \approx y_{0}\left( x\right) +y_{1}\left( x\right)
+y_{2}\left( x\right) +y_{3}\left( x\right) +y_{4}\left( x\right) =1-x^{2}+%
\frac{3x^{4}}{2}-\frac{x^{6}}{2}+\frac{x^{8}}{8}....  \label{34n}
\end{equation}

On the other hand by series expanding the exact solution (\ref{ex1}) we obtain
\begin{equation}
y\left( x\right) =3e^{-x^{2}}+2\left( x^{2}-1\right) =1-x^{2}+\frac{3x^{4}}{2%
}-\frac{x^{6}}{2}+\frac{x^{8}}{8}...,  \label{ns2}
\end{equation}

Clearly, the solution (\ref{34n}) obtained by the Adomian Decomposition Method is
identical to the exact solution (\ref{ns2}).

\subsection{Bernoulli differential equation $\frac{dy}{dx}+P\left( x\right)
y=Q\left( x\right) y^{n}$}

The Adomian Decomposition Method is very powerful for solving nonlinear ordinary
differential equations. Consider that the differential equation takes the
Bernoulli equation form
\begin{equation}
\frac{dy}{dx}+P\left( x\right) y=Q\left( x\right) y^{n},  \label{a1n}
\end{equation}%
where $P\left( x\right) $ and $Q\left( x\right) $ are arbitrary function of $%
x$, and $n$ is an arbitrary constant. Assume that the solution of Eq. (\ref%
{a1n}) is given by the power series form%
\begin{equation}
y\left( x\right) =\sum_{n=0}^{\infty }y_{n}\left( x\right) .  \label{a2n}
\end{equation}%
The nonlinear term $y^{n}$ can be decomposed in terms of the Adomian polynomials $%
A_{n}\left( x\right) $, given by
\begin{equation}
y^{n}\left( x\right) =\sum_{n=0}^{\infty }A_{n}\left( x\right) ,  \label{a3}
\end{equation}.

Generally, for an arbitrary function $f(t,x)$, the Adomian polynomials are defined as \citep{R2}
\be
A_n=\left.\frac{1}{n!}\frac{d^n}{d\epsilon ^n}f\left(t,\sum _{i=0}^{\infty}{\epsilon ^iy_i}\right)\right|_{\epsilon =0}.
\ee

The first four Adomian polynomials can be obtained in the following form,
\be
A_0=f\left(t,y_0\right),
A_1=y_1f'\left(t,y_0\right),
A_2=y_2f'\left(t,y_0\right)+\frac{1}{2}y_1^2f''\left(t,y_0\right),
\ee
\be
A_3=y_3f'\left(t,y_0\right)+y_1y_2f''\left(t,y_0\right)+\frac{1}{6}y_1^3f'''\left(t,y_0\right).
\ee

For the function $y^n$ a few Adomian polynomials are \citep{b7a}
\begin{equation}
A_{0}=y_{0}^{n},  \label{h1}
A_{1}=ny_{1}y_{0}^{n-1},
A_{2}=ny_{2}y_{0}^{n-1}+n\left( n-1\right) \frac{y_{1}^{2}}{2!}y_{0}^{n-2},
\end{equation}%
\begin{equation}
A_{3}=ny_{3}y_{0}^{n-1}+n\left( n-1\right) y_{1}y_{2}y_{0}^{n-2}+n\left(
n-1\right) \left( n-2\right) \frac{y_{1}^{3}}{3!}y_{0}^{n-3}.  \label{h4n}
\end{equation}%
Now integrating Eq. (\ref{a1n}) yields the integral equation%
\begin{equation}
y\left( x\right) =y\left( 0\right) +\int_{0}^{x}\left[ Q\left( x\right)
y^{n}-P\left( x\right) y\right] dx,  \label{a4n}
\end{equation}%
where $y\left( 0\right) $ is the initial condition. Substituting Eqs. (\ref%
{a2n}) and (\ref{a3}) into Eq. (\ref{a4n}) gives the relation%
\begin{equation}
\sum_{n=0}^{\infty }y_{n}\left( x\right) =y\left( 0\right)
+\int_{0}^{x}Q\left( x\right) \sum_{n=0}^{\infty }A_{n}\left( x\right)
dx-\int_{0}^{x}P\left( x\right) \sum_{n=0}^{\infty }y_{n}\left( x\right) dx.
\label{a5n}
\end{equation}%
We rewrite Eq. (\ref{a5n}) in the recursive forms
\begin{equation}
y_{0}\left( x\right) =y\left( 0\right) ,  \label{aan}
\end{equation}%
\begin{equation}
y_{k+1}\left( x\right) =\int_{0}^{x}\left[ Q\left( x\right) A_{k}\left(
x\right) -P\left( x\right) y_{k}\left( x\right) \right] dx.  \label{a6n}
\end{equation}%
From Eqs. (\ref{aan}) and (\ref{a6n}), we obtain the semi-analytical solution of Eq.
(\ref{a1n}), given by
\be
y\left( x\right) =\sum_{n=0}^{\infty }y_{n}\left(
x\right) .
\ee

\subsubsection{Example: $\frac{dy}{dx}-2xy=-4x^{3}y^{2}$}

 Consider now the differential equation
\begin{equation}
\frac{dy}{dx}-2xy=-4x^{3}y^{2},
\end{equation}%
with initial condition $y\left( 0\right) =1$, having the general solution%
\begin{equation}\label{berex}
y\left( x\right) =\frac{1}{3e^{-x^{2}}+2\left( x^{2}-1\right) }.
\end{equation}%
In this case  $P\left( x\right) =-2x$
and $Q\left( x\right) =-4x^{3}$, respectively,  and $n=2$. Next we compute a few Adomian
polynomials for $y^2$,
\begin{equation}
A_{0}=y_{0}^{2},
A_{1}=2y_{1}y_{0},
A_{2}=2y_{2}y_{0}+y_{1}^{2},
A_{3}=2y_{3}y_{0}+2y_{1}y_{2},  \label{a11}
\end{equation}
Hence we obtain
\begin{equation}
y_{0}\left( x\right) =y\left( 0\right) =1,
\end{equation}%
\begin{equation}
y_{k+1}\left( x\right) =\int_{0}^{x}\left[ Q\left( x\right) A_{k}\left(
x\right) -P\left( x\right) y_{k}\left( x\right) \right] dx.
\end{equation}%
Eq. (\ref{a6n}) can be written recursively for $k=0,1,2,3$ in the decomposed
solutions%
\begin{equation}
y_{1}\left( x\right) =\int_{0}^{x}\left[ -4x^{3}A_{0}\left( x\right)
+2xy_{0}\left( x\right) \right] dx=x^{2}-x^{4},
\end{equation}%
\begin{equation}
y_{2}\left( x\right) =\int_{0}^{x}\left[ -4x^{3}A_{1}\left( x\right)
+2xy_{1}\left( x\right) \right] dx=\frac{x^{4}}{2}-\frac{5x^{6}}{3}+x^{8},
\end{equation}%
\begin{equation}
y_{3}\left( x\right) =\int_{0}^{x}\left[ -4x^{3}A_{2}\left( x\right)
+2xy_{2}\left( x\right) \right] dx=\frac{x^{6}}{6}-\frac{17x^{8}}{12}+%
\frac{7x^{10}}{3}-x^{12},
\end{equation}%
\begin{equation}
y_{4}\left( x\right) =\int_{0}^{x}\left[ -4x^{3}A_{3}\left( x\right)
+2xy_{3}\left( x\right) \right] dx=\frac{x^{8}}{24}-\frac{49x^{10}}{60}+%
\frac{25x^{12}}{9}-3x^{14}+x^{16},
\end{equation}%
\begin{equation}
y\left( x\right) \approx y_{0}\left( x\right) +y_{1}\left( x\right)
+y_{2}\left( x\right) +y_{3}\left( x\right) +y_{4}\left( x\right) =1+x^{2}-%
\frac{x^{4}}{2}-\frac{3x^{6}}{2}-\frac{3x^{8}}{8}....  \label{M1}
\end{equation}

On the other hand from the exact solution (\ref{berex}) it is easy to obtain%
\begin{equation}
y\left( x\right) =\frac{1}{3e^{-x^{2}}+2\left( x^{2}-1\right) }=1+x^{2}-%
\frac{x^{4}}{2}-\frac{3x^{6}}{2}-\frac{3x^{8}}{8}....  \label{M2}
\end{equation}%
Clearly again, the solution (\ref{M1}) obtained by the Adomian Decomposition
Method is identical to the exact solution (\ref{M2}).

\subsection{Riccati differential equation $\frac{dy}{dx}=P\left( x\right)
+Q\left( x\right) y^{2}$}

The reduced Riccati differential equation is given by \citep{kamke}
\begin{equation}
\frac{dy}{dx}=P\left( x\right) +Q\left( x\right) y^{2},  \label{r1}
\end{equation}
 where $P\left( x\right) $ and $Q\left( x\right) $ are two arbitrary
functions of $x$, and which must be considered together with the initial condition  $y_0=y\left( 0\right) $. Integrating Eq. (\ref{r1}) yields the equivalent integral equation%
\begin{equation}
y\left( x\right) =y\left( 0\right) +\int_{0}^{x}P\left( x\right)
dx+\int_{0}^{x}Q\left( x\right) y^{2}dx,  \label{r2}
\end{equation}

 Substituting $y\left(
x\right) =\sum_{n=0}^{\infty }y_{n}\left( x\right) $ and $%
y^{2}=\sum_{n=0}^{\infty }A_{n}\left( x\right) $ into Eq. (\ref{r2}) gives
the relation%
\begin{equation}
\sum_{n=0}^{\infty }y_{n}\left( x\right) =y\left( 0\right)
+\int_{0}^{x}P\left( x\right) dx+\int_{0}^{x}Q\left( x\right)
\sum_{n=0}^{\infty }A_{n}dx.  \label{r3}
\end{equation}

Next we rewrite Eq. (\ref{r3}) in the recursive forms%
\begin{equation}
y_{0}\left( x\right) =y\left( 0\right) +\int_{0}^{x}P\left( x\right) dx,
\label{r5}
\end{equation}%
\begin{equation}
y_{k+1}\left( x\right) =\int_{0}^{x}Q\left( x\right) A_{k}\left( x\right) dx.
\label{r6}
\end{equation}%
From Eqs. (\ref{r5}) and (\ref{r6}), we obtain the semi-analytical solution of Eq.
(\ref{r1}), given by $y\left( x\right) =\sum_{n=0}^{\infty }=y_{n}\left(
x\right) $.

\subsubsection{Example: $\frac{dy}{dx}=2e^{x}-e^{-x}y^{2}$}

We consider a particular  Riccati equation that has the form%
\begin{equation}
\frac{dy}{dx}=2e^{x}-e^{-x}y^{2},
\end{equation}
and which must be solved together with the initial condition $y\left( 0\right) =2$.
The general solution of the equation is given by%
\begin{equation}
y\left( x\right) =e^{x}\left( 1-\frac{3}{1-4e^{3x}}\right).
\end{equation}
The semi - analytic solution of this particular Riccati equation can be obtained as
\begin{equation}
y_{0}\left( x\right) =y\left( 0\right) +\int_{0}^{x}P\left( x\right)
dx=2+2\int_{0}^{x}e^{x}dx=2e^{x},  \label{d0}
\end{equation}%
\begin{equation}
y_{k+1}\left( x\right) =\int_{0}^{x}Q\left( x\right) A_{k}\left( x\right)
dx=-\int_{0}^{x}e^{-x}A_{k}dx.  \label{d1}
\end{equation}

In view of Eqs. (\ref{d0}), (\ref{d1}), and (\ref{a11}), we have
\bea
y_{1}\left( x\right) &=&\int_{0}^{x}Q\left( x\right) A_{0}\left( x\right)
dx=-\int_{0}^{x}e^{-x}A_{0}dx=-4x-2x^{2}-\frac{2x^{3}}{3}-\frac{x^{4}}{6}-\nonumber\\
&&\frac{x^{5}}{30}-\frac{x^{6}}{180}-\frac{x^{7}}{1260}...,
\eea
\bea
y_{2}\left( x\right) &=&\int_{0}^{x}Q\left( x\right) A_{1}\left( x\right)
dx=-\int_{0}^{x}e^{-x}A_{1}dx=8x^{2}+\frac{8x^{3}}{3}+\frac{2x^{4}}{3}+\frac{%
2x^{5}}{15}+\nonumber\\
&&\frac{x^{6}}{45}+\frac{x^{7}}{315}...,
\eea
\bea
y_{3}\left( x\right) &=&\int_{0}^{x}Q\left( x\right) A_{2}\left( x\right)
dx=-\int_{0}^{x}e^{-x}A_{2}dx=-16x^{3}-\frac{8x^{4}}{3}-\frac{4x^{5}}{5}-\nonumber\\
&&\frac{4x^{6}}{45}-\frac{2x^{7}}{105}...,
\eea
\begin{equation}
y_{4}\left( x\right) =\int_{0}^{x}Q\left( x\right) A_{3}\left( x\right)
dx=-\int_{0}^{x}e^{-x}A_{3}dx=32x^{4}+\frac{64x^{6}}{45}-\frac{16x^{7}}{315}%
...,
\end{equation}%
\begin{equation}
y\left( x\right) \approx y_{0}\left( x\right) +y_{1}\left( x\right)
+y_{2}\left( x\right) +y_{3}\left( x\right) +y_{4}\left( x\right)
=2-2x+7x^{2}-\frac{41x^{3}}{3}+\frac{359x^{4}}{12}....  \label{v1}
\end{equation}%
From the exact solution by series expansion it is easy to obtain%
\begin{equation}
y\left( x\right) =e^{x}\left( 1-\frac{3}{1-4e^{3x}}\right) =2-2x+7x^{2}-%
\frac{41x^{3}}{3}+\frac{359x^{4}}{12}....  \label{v2}
\end{equation}%
Clearly, the solution (\ref{v1}) obtained by the Adomian decomposition
method is identical to the exact solution (\ref{v2}).

\subsection{ Abel differential equation $\frac{dy}{dx}%
=M\left( x\right) +S\left( x\right) y+R\left( x\right) y^{2}+T\left(
x\right) y^{3}$}

The first kind Abel differential equation takes the form \citep{kamke}
\begin{equation}
\frac{dy}{dx}=M\left( x\right) +S\left( x\right) y+R\left( x\right)
y^{2}+T\left( x\right) y^{3}.  \label{m1}
\end{equation}%
Integrating Eq. (\ref{m1}) yields the relation%
\begin{equation}
y\left( x\right) =y\left( 0\right) +\int_{0}^{x}M\left( x\right)
dx+\int_{0}^{x}\left[ S\left( x\right) y+R\left( x\right) y^{2}+T\left(
x\right) y^{3}\right] dx.  \label{m2}
\end{equation}

Inserting $y\left( x\right) =\sum_{n=0}^{\infty }y_{n}\left( x\right) $, $%
y^{2}=\sum_{n=0}^{\infty }A_{n}\left( x\right) $ and $y^{3}=\sum_{n=0}^{%
\infty }B_{n}\left( x\right) $ into Eq. (\ref{m2}) gives the relation
\bea
&&\sum_{n=0}^{\infty }y_{n}\left( x\right) =y\left( 0\right)
+\nonumber\\
&&\int_{0}^{x}M\left( x\right) dx+\int_{0}^{x}\left[ S\left( x\right)
\sum_{n=0}^{\infty }y_{n}\left( x\right) +R\left( x\right)
\sum_{n=0}^{\infty }A_{n}\left( x\right) +T\left( x\right)
\sum_{n=0}^{\infty }B_{n}\left( x\right) \right] dx.\nonumber\\
\eea
Then we have

\begin{equation}
y_{0}\left( x\right) =y\left( 0\right) +\int_{0}^{x}M\left( x\right) dx,
\label{x1}
\end{equation}%
\begin{equation}
y_{k+1}\left( x\right) =\int_{0}^{x}\left[ S\left( x\right) y_{k}\left(
x\right) +R\left( x\right) A_{k}\left( x\right) +T\left( x\right)
B_{k}\left( x\right) \right] dx.  \label{x2}
\end{equation}%
From Eqs. (\ref{x1}) and (\ref{x2}), we can obtain the semi-analytical solution
of the Abel Eq. (\ref{m1}) as given by $y\left( x\right) =\sum_{n=0}^{\infty
}y_{n}\left( x\right) $.

\subsubsection{Example: $\frac{dy}{dx}=x+3xy+3xy^{2}+xy^{3}$}

We consider now a first kind Abel equation that has the form%
\begin{equation}
\frac{dy}{dx}=x+3xy+3xy^{2}+xy^{3}=x\left( 1+y\right) ^{3},
\end{equation}
which should be solved with initial condition $y\left( 0\right) =0$, or $y\left( 0\right) =-2$.
Its general solution is given by%
\begin{equation}
y\left( x\right) =-1\pm \frac{1}{\sqrt{1-x^{2}}}.
\end{equation}

Now $%
M\left( x\right) =T\left( x\right) =x$ and $S\left( x\right) =R\left(
x\right) =3x$, and a few Adomian polynomials of $y^3$  are%
\begin{equation}
B_{0}=y_{0}^{3},
B_{1}=3y_{1}y_{0}^{2},
B_{2}=3y_{2}y_{0}^{2}+3y_{1}^{2}y_{0},
B_{3}=3y_{3}y_{0}^{2}+6y_{1}y_{2}y_{0}+y_{1}^{3}.  \label{a15n}
\end{equation}

With the help of Eqs. (\ref{x1}), (\ref{x2}), (\ref{a11}), and (\ref%
{a15n}), by taking $y\left( 0\right) =0$, we obtain%
\begin{equation}
y_{0}\left( x\right) =y\left( 0\right) +\int_{0}^{x}M\left( x\right) dx=%
\frac{x^{2}}{2},
\end{equation}
\begin{equation}
y_{1}\left( x\right) =\int_{0}^{x}\left[ 3xy_{0}\left( x\right)
+3xA_{0}\left( x\right) +xB_{0}\left( x\right) \right] dx=\frac{3x^{4}}{8}+%
\frac{x^{6}}{8}+\frac{x^{8}}{64},
\end{equation}%
\bea
y_{2}\left( x\right) &=&\int_{0}^{x}\left[ 3xy_{1}\left( x\right)
+3xA_{1}\left( x\right) +xB_{1}\left( x\right) \right] dx=\frac{3x^{6}}{16}+%
\frac{3x^{8}}{16}+\frac{9x^{10}}{128}+\nonumber\\
&&\frac{3x^{12}}{256}+\frac{3x^{14}}{3584%
},
\eea
\begin{equation}
y_{3}\left( x\right) =\int_{0}^{x}\left[ 3xy_{2}\left( x\right)
+3xA_{2}\left( x\right) +xB_{2}\left( x\right) \right] dx=\frac{9x^{8}}{128}+%
\frac{99x^{10}}{640}+\frac{15x^{12}}{128}...,
\end{equation}%
\begin{equation}
y_{4}\left( x\right) =\int_{0}^{x}\left[ 3xy_{3}\left( x\right)
+3xA_{3}\left( x\right) +xB_{3}\left( x\right) \right] dx=\frac{27x^{10}}{%
1280}+\frac{117x^{12}}{1280}+\frac{2169x^{14}}{17920}...,
\end{equation}%
\begin{equation}
y\left( x\right) \approx y_{0}\left( x\right) +y_{1}\left( x\right)
+y_{2}\left( x\right) +y_{3}\left( x\right) +y_{4}\left( x\right) =\frac{%
x^{2}}{2}+\frac{3x^{4}}{8}+\frac{5x^{6}}{16}+\frac{35x^{8}}{128}+\frac{%
63x^{10}}{256}....  \label{b1}
\end{equation}%
From the exact solution it is easy to obtain%
\begin{equation}
y\left( x\right) =-1+\frac{1}{\sqrt{1-x^{2}}}=\frac{x^{2}}{2}+\frac{3x^{4}}{8%
}+\frac{5x^{6}}{16}+\frac{35x^{8}}{128}+\frac{63x^{10}}{256}....  \label{b2}
\end{equation}%
It immediately follows that the solution (\ref{b1}) obtained by the Adomian Decomposition
Method is identical to the exact solution (\ref{b2}).

\section{Solving second order ordinary differential equations via Adomian decomposition
method}\label{sect3}

Consider a second order non-linear  differential equation that takes the form%
\begin{equation}
\frac{d^{2}y}{dx^{2}}+f\left( x\right) \frac{dy}{dx}+s\left( x\right)
y+g\left( x\right) y^{n}=k\left( x\right) ,  \label{w1}
\end{equation}
and which must be solved together with the initial conditions  $y\left( 0\right) $ and $y^{\prime }\left( 0\right) $, respectively, where $f\left( x\right) $, $s\left( x\right) $, $g\left( x\right) $ and $%
k\left( x\right) $ are  arbitrary function of $x$, and $n$ is a
constant.  We define the  integral operator $L^{-1}$ as
\begin{equation}
L^{-1}\left( .\right) =\int_{0}^{x}e^{-\int f\left( x\right)
dx}\int_{0}^{x}e^{\int f\left( x\right) dx}\left( .\right) dxdx.
\end{equation}%

We consider first the action of the integral operator $L^{-1}$ on the first two terms of the equation, which gives
\begin{eqnarray}
L^{-1}\left[ \frac{d^{2}y}{dx^{2}}+f\left( x\right) \frac{dy}{dx}\right]
&=&\int_{0}^{x}e^{-\int f\left( x\right) dx}\int_{0}^{x}e^{\int f\left(
x\right) dx}\left[ \frac{d^{2}y}{dx^{2}}+f\left( x\right) \frac{dy}{dx}%
\right] dxdx \nonumber\\
&=&\int_{0}^{x}e^{-\int f\left( x\right) dx}\left( \int_{0}^{x}e^{\int
f\left( x\right) dx}dy^{\prime }+\int_{0}^{x}e^{\int f\left( x\right)
dx}fy^{\prime }dx\right) dx \nonumber\\
&=&\int_{0}^{x}e^{-\int f\left( x\right) dx}\left\{ \left[ e^{\int f\left(
x\right) dx}y^{\prime }\right] _{0}^{x}\right\} dx \nonumber\\
&=&\int_{0}^{x}y^{\prime }dx-\left[ e^{\int f\left( x\right) dx}\right]
_{x=0}y^{\prime }\left( 0\right) \int_{0}^{x}e^{-\int f\left( x\right) dx}dx
\nonumber\\
&=&y\left( x\right) -y\left( 0\right) -y^{\prime }\left( 0\right) \left[
e^{\int f\left( x\right) dx}\right] _{x=0}\int_{0}^{x}e^{-\int f\left(
x\right) dx}dx.
\end{eqnarray}%

Then we have
\begin{eqnarray}
L^{-1}\left[ \frac{d^{2}y}{dx^{2}}+f\left( x\right) \frac{dy}{dx}\right]
&=&L^{-1}\left[ k\left( x\right) -s\left( x\right) y-g\left( x\right) y^{n}%
\right],
\eea
\bea
y\left( x\right) &=&\phi \left( x\right) +\int_{0}^{x}e^{-\int f\left(
x\right) dx}\left\{ \int_{0}^{x}e^{\int f\left( x\right) dx}\left[ k\left(
x\right) -s\left( x\right) y-g\left( x\right) y^{n}\right] dx\right\} dx,\nonumber\\
\end{eqnarray}%
where we have denoted $\phi \left( x\right) $ as
\begin{equation}
\phi \left( x\right) =y\left( 0\right) +y^{\prime }\left( 0\right) \left[
e^{\int f\left( x\right) dx}\right] _{x=0}\int_{0}^{x}e^{-\int f\left(
x\right) dx}dx.
\end{equation}
Hence we obtain
\begin{eqnarray}
\sum_{n=0}^{\infty }y_{n}\left( x\right) &=&\phi \left( x\right)
+\int_{0}^{x}e^{-\int f\left( x\right) dx}\left[ \int_{0}^{x}e^{\int f\left(
x\right) dx}k\left( x\right) dx\right] dx-  \notag \\
&&\int_{0}^{x}e^{-\int f\left( x\right) dx}\left\{ \int_{0}^{x}e^{\int
f\left( x\right) dx}\left[ s\left( x\right) \sum_{n=0}^{\infty }y_{n}\left(
x\right) +g\left( x\right) \sum_{n=0}^{\infty }A_{n}\left( x\right) \right]
dx\right\} dx. \nonumber\\
\end{eqnarray}%
Then for the solution of the second order nonlinear differential equation we have
\begin{equation}\label{g1}
y_{0}\left( x\right) =\phi \left( x\right) +\int_{0}^{x}e^{-\int f\left(
x\right) dx}\left[ \int_{0}^{x}e^{\int f\left( x\right) dx}k\left( x\right)
dx\right] dx,
\end{equation}%
\begin{equation}
y_{k+1}\left( x\right) =-\int_{0}^{x}e^{-\int f\left( x\right) dx}\left[
\int_{0}^{x}e^{\int f\left( x\right) dx}\left[ s\left( x\right) y_{k}\left(
x\right) +g\left( x\right) A_{k}\left( x\right) \right] dx\right] dx.
\label{g2}
\end{equation}%
From Eqs. (\ref{h1})-(\ref{h4n}), and (\ref{g1}), (\ref{g2}), we  obtain the
semi-analytical solution of Eq. (\ref{w1}), given by $y=\sum_{n=0}^{\infty
}y_{n}\left( x\right) $.

\subsection{Example: $\frac{d^{2}y}{dx^{2}}+4\frac{dy}{dx}+3y=3$}

As an example of the application of the ADM we consider a particular second order differential equation that takes the form%
\begin{equation}
\frac{d^{2}y}{dx^{2}}+4\frac{dy}{dx}+3y=3,  \label{ss1}
\end{equation}%
which must be considered together with the initial conditions $y\left( 0\right) =1$ and $y^{\prime }\left(
0\right) =2$. The general solution of the equation is given by
\begin{equation}
y\left( x\right) =-e^{-3x}+e^{-x}+1,
\end{equation}%
 From the equation we easily obtain  $e^{\int f\left( x\right) dx}=e^{\int
4dx}=e^{4x}$, $f\left( x\right) =4,$ $g\left( x\right) =0$, $k\left(
x\right) =3$ and $s\left( x\right) =3$, respectively. Then we have
\begin{eqnarray}
\sum_{n=0}^{\infty }y_{n}\left( x\right) &=&\phi \left( x\right)
+\int_{0}^{x}e^{-\int f\left( x\right) dx}\left[ \int_{0}^{x}e^{\int f\left(
x\right) dx}k\left( x\right) dx\right] dx-  \notag \\
&&\int_{0}^{x}e^{-\int f\left( x\right) dx}\left[ \int_{0}^{x}e^{\int
f\left( x\right) dx}s\left( x\right) \sum_{n=0}^{\infty }y_{n}\left(
x\right) dx\right] dx.  \label{bb}
\end{eqnarray}%
We rewrite Eq. (\ref{bb}) in the recursive forms

\begin{eqnarray}
y_{0}\left( x\right) &=&y\left( 0\right) +y^{\prime }\left( 0\right) \left[
e^{\int f\left( x\right) dx}\right] _{x=0}\int_{0}^{x}e^{-\int f\left(
x\right) dx}dx+ \\
&&\int_{0}^{x}e^{-\int f\left( x\right) dx}\left[ \int_{0}^{x}e^{\int
f\left( x\right) dx}k\left( x\right) dx\right] dx \\
&=&\frac{21}{16}+\frac{3}{4}x-\frac{5}{16}e^{-4x},
\end{eqnarray}%
and%
\begin{equation}
y_{k+1}\left( x\right) =-3\int_{0}^{x}e^{-4x}\left[ \int_{0}^{x}e^{4x}y_{k}%
\left( x\right) dx\right] dx.
\end{equation}%
Hence we obtain
\begin{eqnarray}
y_{1}\left( x\right) &=&-3\int_{0}^{x}e^{-4x}\left[ \int_{0}^{x}e^{4x}y_{0}%
\left( x\right) dx\right] dx \\
&=&-\frac{3}{2}x^{2}+x^{3}-\frac{3x^{4}}{8}-\frac{x^{5}}{5}+\frac{7x^{6}}{15}%
-\frac{16x^{7}}{35}+\frac{34x^{8}}{105}...,
\end{eqnarray}%
\begin{eqnarray}
y_{2}\left( x\right) &=&-3\int_{0}^{x}e^{-4x}\left[ \int_{0}^{x}e^{4x}y_{1}%
\left( x\right) dx\right] dx \\
&=&\frac{3x^{4}}{8}-\frac{9x^{5}}{20}+\frac{27x^{6}}{80}-\frac{5x^{7}}{28}+%
\frac{9x^{8}}{140}...,
\end{eqnarray}%
\begin{eqnarray}
y_{3}\left( x\right) &=&-3\int_{0}^{x}e^{-4x}\left[ \int_{0}^{x}e^{4x}y_{2}%
\left( x\right) dx\right] dx \\
&=&-\frac{3x^{6}}{80}+\frac{3x^{7}}{56}-\frac{201x^{8}}{4480}+\frac{23x^{9}}{%
840}-\frac{11x^{10}}{840},
\end{eqnarray}%
\begin{eqnarray}
y_{4}\left( x\right) &=&-3\int_{0}^{x}e^{-4x}\left[ \int_{0}^{x}e^{4x}y_{3}%
\left( x\right) dx\right] dx \\
&=&\frac{9x^{8}}{4480}-\frac{x^{9}}{320}+\frac{123x^{10}}{44800}....
\end{eqnarray}%
The semi-analytical solution of Eq. (\ref{ss1}) is given by
\begin{eqnarray}\label{t1}
y\left( x\right) &\approx &y_{0}\left( x\right) +y_{1}\left( x\right)
+y_{2}\left( x\right) +y_{3}\left( x\right) +...  \nonumber\\
&=&1+2x-4x^{2}+\frac{13x^{3}}{3}-\frac{10x^{4}}{3}+\frac{121x^{5}}{60}-\frac{%
91x^{6}}{90}+\frac{1093x^{7}}{2520}....
\end{eqnarray}%
On the other hand from the exact solution it is easy to obtain
\bea \label{t2}
y\left( x\right) &=&-e^{-3x}+e^{-x}+1=1+2x-4x^{2}+\frac{13x^{3}}{3}-\frac{%
10x^{4}}{3}+\frac{121x^{5}}{60}-\nonumber\\
&&\frac{91x^{6}}{90}+\frac{1093x^{7}}{2520}+%
....
\eea

As one can easily see, the solution (\ref{t1}) of the second order differential Eq. (\ref%
{ss1}) obtained by the Adomian decomposition method is identical to the
exact solution (\ref{t2}).

\subsection{Example: $\frac{d^{2}y}{dx^{2}}+\frac{dy}{dx}+e^{x}y+e^{x}y^{2}=e^{x}$}

As a second example of the application of the ADM for solving a nonlinear differential equation we consider that the differential equation takes the form%
\begin{equation}
\frac{d^{2}y}{dx^{2}}+\frac{dy}{dx}+e^{x}y+e^{x}y^{2}=e^{x},  \label{j1}
\end{equation}%
and it must be solved together with the initial condition $y\left( 0\right) =1$ and $y^{\prime }\left(
0\right) =2$, respectively. From the equation we obtain easily  $e^{\int f\left( x\right) dx}=e^{\int
dx}=e^{x} $, $f\left( x\right) =1,$ $g\left( x\right) =k\left( x\right)
=s\left( x\right) =e^{x}$. Hence we immediately find
\begin{eqnarray}
y_{0}\left( x\right) &=&y\left( 0\right) +y^{\prime }\left( 0\right) \left[
e^{\int f\left( x\right) dx}\right] _{x=0}\int_{0}^{x}e^{-\int f\left(
x\right) dx}dx+  \notag \\
&&\int_{0}^{x}e^{-\int f\left( x\right) dx}\left[ \int_{0}^{x}e^{\int
f\left( x\right) dx}k\left( x\right) dx\right] dx \\
&=&1+2\int_{0}^{x}e^{-x}dx+\int_{0}^{x}e^{-x}\left[ \int_{0}^{x}e^{2x}dx%
\right] dx  \notag \\
&=&2-\frac{3}{2}e^{-x}+\frac{1}{2}e^{x}.  \label{c1}
\end{eqnarray}%
and
\begin{eqnarray}
y_{k+1}\left( x\right) &=&-\int_{0}^{x}e^{-\int f\left( x\right) dx}\left[
\int_{0}^{x}e^{\int f\left( x\right) dx}\left[ s\left( x\right) y_{k}\left(
x\right) +g\left( x\right) A_{k}\left( x\right) \right] dx\right] dx,  \notag
\\
y_{k+1}\left( x\right) &=&-\int_{0}^{x}e^{-x}\left[ \int_{0}^{x}e^{2x}\left[
y_{k}\left( x\right) +A_{k}\left( x\right) \right] dx\right] dx,  \label{c2}
\end{eqnarray}%
From Eqs. (\ref{a11}), (\ref{c1}), and (\ref{c2}), we obtain
\begin{eqnarray}
y_{1}\left( x\right) &=&-\int_{0}^{x}e^{-x}\left[ \int_{0}^{x}e^{2x}\left[
y_{0}\left( x\right) +A_{0}\left( x\right) \right] dx\right] dx \\
&=&-x^{2}-x^{3}-\frac{13x^{4}}{24}-\frac{2x^{5}}{15}-\frac{17x^{6}}{240}-%
\frac{23x^{7}}{1260}-\frac{257x^{8}}{40320}-\nonumber\\
&&\frac{17x^{9}}{10080}-\frac{67x^{10}}{145152}...,
\end{eqnarray}%
\begin{eqnarray}
y_{2}\left( x\right) &=&-\int_{0}^{x}e^{-x}\left[ \int_{0}^{x}e^{2x}\left[
y_{1}\left( x\right) +A_{1}\left( x\right) \right] dx\right] dx \\
&=&\frac{x^{4}}{4}+\frac{9x^{5}}{20}+\frac{29x^{6}}{80}+\frac{521x^{7}}{2520}%
+\frac{233x^{8}}{2520}+\frac{155x^{9}}{4032}+\frac{17293x^{10}}{1209600}...,
\end{eqnarray}%
\begin{eqnarray}
y_{3}\left( x\right) &=&-\int_{0}^{x}e^{-x}\left[ \int_{0}^{x}e^{2x}\left[
y_{2}\left( x\right) +A_{2}\left( x\right) \right] dx\right] dx \\
&=&-\frac{7x^{6}}{120}-\frac{23x^{7}}{168}-\frac{719x^{8}}{4480}-\frac{%
3797x^{9}}{30240}-\frac{5143x^{10}}{67200}...,
\end{eqnarray}%
\bea
y_{4}\left( x\right) &=&-\int_{0}^{x}e^{-x}\left[ \int_{0}^{x}e^{2x}\left[
y_{3}\left( x\right) +A_{3}\left( x\right) \right] dx\right] dx=\nonumber\\
&&\frac{27x^{8}%
}{2240}+\frac{2203x^{9}}{60480}+\frac{3149x^{10}}{57600}....
\eea

Hence the semi-analytical solution of Eq. (\ref{j1}) is given by
\begin{eqnarray}
y\left( x\right) &\approx &y_{0}\left( x\right) +y_{1}\left( x\right)
+y_{2}\left( x\right) +y_{3}\left( x\right) +...  \notag \\
&=&1+2x-\frac{3x^{2}}{2}-\frac{2x^{3}}{3}-\frac{x^{4}}{3}+\frac{x^{5}}{3}+%
\frac{167x^{6}}{720}+\frac{131x^{7}}{2520}-\frac{503x^{8}}{8064}-\frac{%
9503x^{9}}{181440}-\nonumber\\
&&\frac{7283x^{10}}{907200}....  \label{jj}
\end{eqnarray}

\section{The fifth order ordinary differential equation via the Adomian Decomposition Method}\label{sect4}

Consider the following fifth order ordinary differential equation, which  takes the form
\begin{equation}
\frac{d^{5}y}{dx^{5}}+a_{4}\frac{d^{4}y}{dx^{4}}+a_{3}\frac{d^{3}y}{dx^{3}}%
+a_{2}\frac{d^{2}y}{dx^{2}}+a_{1}\frac{dy}{dx}+a_{0}y=0,  \label{51}
\end{equation}
where $a_{i}$, $i=0,1,2,3,4$ are constants. Eq.~(\ref{51}) should be integrated with the initial conditions $y(0)=y_0$, $y'(0)=y_{01}$, $y''(0)=y_{02}$, $y'''(0)=y_{03}$, and $y^{(iv)}(0)=y_{04}$, respectively.  Now applying the 5 fold
integral operator $
L^{-1}$, defined as
\be
L^{-1}\left( .\right)
=\int_{0}^{x}\int_{0}^{x}\int_{0}^{x}\int_{0}^{x}\int_{0}^{x}\left( .\right)
dxdxdxdxdx,
\ee
 to Eq. (\ref{51}), yields the relation%
\begin{equation}
\int_{0}^{x}\int_{0}^{x}\int_{0}^{x}\int_{0}^{x}\int_{0}^{x}\left( \frac{%
d^{5}y}{dx^{5}}+a_{4}\frac{d^{4}y}{dx^{4}}+a_{3}\frac{d^{3}y}{dx^{3}}+a_{2}%
\frac{d^{2}y}{dx^{2}}+a_{1}\frac{dy}{dx}+a_{0}y\right) d^5x=0.
\label{52}
\end{equation}

From Eq. (\ref{52}), we obtain%
\begin{eqnarray}
y\left( x\right) &=&y\left( 0\right) +\left[ y^{\prime }\left( 0\right)
+a_{4}y\left( 0\right) \right] x+\left[ y^{\prime \prime }\left( 0\right)
+a_{4}y^{\prime }\left( 0\right) +a_{3}y\left( 0\right) \right] \frac{x^{2}}{%
2}+\nonumber\\
&&\left[ y^{\prime \prime \prime }\left( 0\right) +a_{4}y^{\prime \prime
}\left( 0\right) +a_{3}y^{\prime }\left( 0\right) +a_{2}y\left( 0\right) %
\right] \frac{x^{3}}{6}+  \notag \\
&&\left[ y^{\prime \prime \prime \prime }\left( 0\right) +a_{4}y^{\prime
\prime \prime }\left( 0\right) +a_{3}y^{\prime \prime }\left( 0\right)
+a_{2}y^{\prime }\left( 0\right) +a_{1}y\left( 0\right) \right] \frac{x^{4}}{%
24}-\nonumber\\
&&a_{4}\int_{0}^{x}ydx-a_{3}\int_{0}^{x}\int_{0}^{x}ydxdx-  \notag \\
&&a_{2}\int_{0}^{x}\int_{0}^{x}\int_{0}^{x}ydxdxdx-a_{1}\int_{0}^{x}%
\int_{0}^{x}\int_{0}^{x}\int_{0}^{x}ydxdxdxdx-\nonumber\\
&&a_{0}\int_{0}^{x}\int_{0}^{x}%
\int_{0}^{x}\int_{0}^{x}\int_{0}^{x}ydxdxdxdxdx.  \label{53}
\end{eqnarray}

Substituting $y\left( x\right) =\sum_{n=0}^{\infty }y_{n}\left( x\right) $
into Eq. (\ref{53}) yields
\begin{eqnarray}
\sum_{n=0}^{\infty }y_{n}\left( x\right) &=&y\left( 0\right) +\left[
y^{\prime }\left( 0\right) +a_{4}y\left( 0\right) \right] x+\left[ y^{\prime
\prime }\left( 0\right) +a_{4}y^{\prime }\left( 0\right) +a_{3}y\left(
0\right) \right] \frac{x^{2}}{2}+\nonumber\\
&&\left[ y^{\prime \prime \prime }\left(
0\right) +a_{4}y^{\prime \prime }\left( 0\right) +a_{3}y^{\prime }\left(
0\right) +a_{2}y\left( 0\right) \right] \frac{x^{3}}{6}+  \notag \\
&&\left[ y^{\prime \prime \prime \prime }\left( 0\right) +a_{4}y^{\prime
\prime \prime }\left( 0\right) +a_{3}y^{\prime \prime }\left( 0\right)
+a_{2}y^{\prime }\left( 0\right) +a_{1}y\left( 0\right) \right] \frac{x^{4}}{%
24}-\nonumber\\
&&a_{4}\int_{0}^{x}\sum_{n=0}^{\infty }y_{n}\left( x\right) dx-
a_{3}\int_{0}^{x}\int_{0}^{x}\sum_{n=0}^{\infty }y_{n}\left( x\right)
dxdx-\nonumber\\
&&a_{2}\int_{0}^{x}\int_{0}^{x}\int_{0}^{x}\sum_{n=0}^{\infty
}y_{n}\left( x\right) dxdxdx-  \nonumber\\
&&a_{1}\int_{0}^{x}\int_{0}^{x}\int_{0}^{x}\int_{0}^{x}\sum_{n=0}^{\infty
}y_{n}\left( x\right)
dxdxdxdx-\nonumber\\
&&a_{0}\int_{0}^{x}\int_{0}^{x}\int_{0}^{x}\int_{0}^{x}\int_{0}^{x}%
\sum_{n=0}^{\infty }y_{n}\left( x\right) dxdxdxdxdx.  \label{54}
\end{eqnarray}

We rewrite Eq. (\ref{54}) in the recursive forms%
\begin{eqnarray}
y_{0}\left( x\right) &=&y\left( 0\right) +\left[ y^{\prime }\left( 0\right)
+a_{4}y\left( 0\right) \right] x+\left[ y^{\prime \prime }\left( 0\right)
+a_{4}y^{\prime }\left( 0\right) +a_{3}y\left( 0\right) \right] \frac{x^{2}}{%
2}+\nonumber\\
&&\left[ y^{\prime \prime \prime }\left( 0\right) +a_{4}y^{\prime \prime
}\left( 0\right) +a_{3}y^{\prime }\left( 0\right) +a_{2}y\left( 0\right) %
\right] \frac{x^{3}}{6}+  \notag \\
&&\left[ y^{\prime \prime \prime \prime }\left( 0\right) +a_{4}y^{\prime
\prime \prime }\left( 0\right) +a_{3}y^{\prime \prime }\left( 0\right)
+a_{2}y^{\prime }\left( 0\right) +a_{1}y\left( 0\right) \right] \frac{x^{4}}{%
24},  \label{55}
\end{eqnarray}
and
\begin{eqnarray}
y_{k+1}\left( x\right) &=&-a_{4}\int_{0}^{x}y_{k}\left( x\right)
dx-a_{3}\int_{0}^{x}\int_{0}^{x}y_{k}\left( x\right)
dxdx-\nonumber\\
&&a_{2}\int_{0}^{x}\int_{0}^{x}\int_{0}^{x}y_{k}\left( x\right) dxdxdx-
\notag \\
&&a_{1}\int_{0}^{x}\int_{0}^{x}\int_{0}^{x}\int_{0}^{x}y_{k}\left( x\right)
dxdxdxdx-\nonumber\\
&&a_{0}\int_{0}^{x}\int_{0}^{x}\int_{0}^{x}\int_{0}^{x}%
\int_{0}^{x}y_{k}\left( x\right) dxdxdxdxdx.  \label{56}
\end{eqnarray}

From Eqs. (\ref{55}) and (\ref{56}), we can obtain the semi-analytical solution
of Eq. (\ref{51}), given by $y\left( x\right) =\sum_{n=0}^{\infty
}y_{n}\left( x\right) $. For the solution of a particular third-order
ordinary differential equation see \citet{PPa}.

\subsection{Example: $\frac{d^{5}y}{dx^{5}}-3\frac{d^{4}y}{dx^{4}}-5\frac{d^{3}y}{%
dx^{3}}+15\frac{d^{2}y}{dx^{2}}+4\frac{dy}{dx}-12y=0$}

In the following we consider a particular fifth order ordinary differential equation that takes the form
\begin{equation}
\frac{d^{5}y}{dx^{5}}-3\frac{d^{4}y}{dx^{4}}-5\frac{d^{3}y}{dx^{3}}+15\frac{%
d^{2}y}{dx^{2}}+4\frac{dy}{dx}-12y=0,  \label{gg1}
\end{equation}
and which should be solved together with the initial conditions  $y\left( 0\right) =1$, $y^{\prime }\left(
0\right) =-1$,$y^{\prime \prime }\left( 0\right) =2$, $y^{\prime \prime
\prime }\left( 0\right) =-2$, $y^{\prime \prime \prime \prime }\left(
0\right) =3$.  The coefficients $a_i$ of the equation are given by $a_{4}=-3$,$a_{3}=-5$,$a_{2}=15$,$a_{1}=4$ and
$a_{0}=-12$, respectively.
The general solution of the equation is given by%
\begin{equation}
y\left( x\right) =-\frac{1}{4}e^{x}+\frac{19}{24}e^{-x}+\frac{1}{3}e^{2x}+%
\frac{1}{5}e^{-2x}-\frac{3}{40}e^{3x}.
\end{equation}

By applying the ADM we have%
\begin{equation}
y_{0}\left( x\right) =1-4x+2x^{3}-\frac{x^{4}}{2},  \label{ee}
\end{equation}%
\begin{eqnarray} \label{ef}
y_{k+1}\left( x\right) &=&3\int_{0}^{x}y_{k}\left( x\right)
dx+5\int_{0}^{x}\int_{0}^{x}y_{k}\left( x\right)
dxdx-\nonumber\\
&&15\int_{0}^{x}\int_{0}^{x}\int_{0}^{x}y_{k}\left( x\right) dxdxdx-
\notag \\
&&4\int_{0}^{x}\int_{0}^{x}\int_{0}^{x}\int_{0}^{x}y_{k}\left( x\right)
dxdxdxdx+\nonumber\\
&&12\int_{0}^{x}\int_{0}^{x}\int_{0}^{x}\int_{0}^{x}\int_{0}^{x}y_{k}%
\left( x\right) dxdxdxdxdx.
\end{eqnarray}%
From Eqs. (\ref{ee}) and (\ref{ef}), we obtain%
\begin{equation}
y_{1}\left( x\right) =3x-\frac{7x^{2}}{2}-\frac{35x^{3}}{6}+\frac{23x^{4}}{6}%
+\frac{13x^{5}}{30}-\frac{2x^{6}}{5}...,
\end{equation}%
\begin{equation}
y_{2}\left( x\right) =\frac{9x^{2}}{2}-x^{3}-\frac{185x^{4}}{24}+\frac{%
97x^{5}}{60}+\frac{241x^{6}}{144}...,
\end{equation}%
\begin{equation}
y_{3}\left( x\right) =\frac{9x^{3}}{2}+\frac{9x^{4}}{8}-6x^{5}-\frac{289x^{6}%
}{720}...,
\end{equation}%
\begin{equation}
y_{4}\left( x\right) =\frac{27x^{4}}{8}+\frac{9x^{5}}{5}-\frac{27x^{6}}{8}%
....
\end{equation}%

Thus we obtain the ADM solution of the equation as
\begin{equation}
y\left( x\right) \approx y_{0}\left( x\right) +y_{1}\left( x\right)
+y_{2}\left( x\right) +y_{3}\left( x\right) +y_{4}\left( x\right) =1-x+x^{2}-%
\frac{x^{3}}{3}+\frac{x^{4}}{8}....  \label{y1}
\end{equation}%
On the other hand from the exact solution it is easy to obtain
\begin{equation}
y\left( x\right) =-\frac{1}{4}e^{x}+\frac{19}{24}e^{-x}+\frac{1}{3}e^{2x}+%
\frac{1}{5}e^{-2x}-\frac{3}{40}e^{3x}=1-x+x^{2}-\frac{x^{3}}{3}+\frac{x^{4}}{%
8}...,  \label{y2}
\end{equation}%
Hence, the solution (\ref{y1}) of the fifth order differential Eq. (\ref%
{gg1}), obtained by the Adomian decomposition method, is identical to the
exact solution (\ref{y2}).

\section{Solving partial differential equations via ADM: the Fisher-Kolmogorov equation}\label{sect5}

The three dimensional Fisher-Kolmogorov equation \citep{Fish, Kolm, TW1}, has many important applications in physics and biology. In particular, it can be used to describe the growth of
glioblastoma \citep{TW2}. The Fisher-Kolmogorov equation is given by
\begin{equation}
\frac{\partial c\left( t,x,y,z\right) }{\partial t}=D\Delta c\left(
t,x,y,z\right) +ac\left( t,x,y,z\right) \left[ 1-\frac{c\left(
t,x,y,z\right) }{N}\right] ,  \label{q1}
\end{equation}
where
\be
\Delta =\frac{\partial ^{2}}{\partial x^{2}}+\frac{\partial ^{2}}{%
\partial y^{2}}+\frac{\partial ^{2}}{\partial z^{2}},
\ee
and $a$, $N$ and $D$
are constants. Eq. (\ref{q1}) must be considered together with the initial condition $%
c\left( 0,x,y,z\right) =c_{0}\left( x,y,z\right) $. From the point of view of the ADM the Fisher-Kolmogorov equation was studied in  \citet{W1} and \citet{Bhal}, respectively.  In order to apply the ADM method we rewrite Eq. (\ref{q1})
as
\begin{equation}
L_{t}c=D\Delta c+F\left( c\right) ,  \label{q2}
\end{equation}%
where $L_{t}=\frac{\partial }{\partial t}$ and $F\left( c\right) =ac\left[ 1-%
\frac{c}{N}\right] $. Now applying the inverse operator $L_{t}^{-1}$ defined
as $L_{t}^{-1}\left( .\right) =\int_{0}^{t}\left( .\right) dt$ to Eq. (\ref%
{q2}), the general solution of Eq. (\ref{q2}) can be obtained formally
as
\begin{equation}
c\left( t,x,y,x\right) =c_{0}\left( x,y,z\right) +DL_{t}^{-1}\Delta c\left(
t,x,y,x\right) +L_{t}^{-1}F\left( c\right) .  \label{q3}
\end{equation}%

According to the Adomian Decomposition Method we look for series solutions
of Eq.~(\ref{q3}) of the form
\begin{equation}
c\left( t,x,y,z\right) =\sum_{n=0}^{\infty }c_{n}\left( t,x,y,z\right)
,F\left( c\right) =\sum_{n=0}^{\infty }A_{n}\left( t,x,y,z\right) ,
\label{q4}
\end{equation}%
where the Adomian polynomials $A_{n}\left( t,x,y,z\right) $ are defined as
\begin{equation}
A_{n}\left( t,x,y,z\right) =\frac{1}{n!}\left[ \frac{d^{n}}{d\lambda ^{n}}%
F\left( c_{\lambda }\right) \right] _{\lambda =0},  \label{5a}
\end{equation}%
where $c_{\lambda }=\sum_{i=0}^{\infty }\lambda ^{i}c_{i}$. The first few
Adomian polynomials are given by

\begin{eqnarray}
A_{0} &=&F\left( c_{0}\right) =ac_{0}\left( 1-\frac{c_{0}}{N}\right) , \\
A_{1} &=&c_{1}F^{\prime }\left( c_{0}\right) =ac_{1}\left( 1-\frac{2c_{0}}{N}%
\right) ,  \label{7} \\
A_{2} &=&c_{2}F^{\prime }\left( c_{0}\right) +\frac{1}{2}c_{1}^{2}F^{\prime
\prime }\left( c_{0}\right) =ac_{2}\left( 1-\frac{2c_{0}}{N}\right) -\frac{a%
}{N}c_{1}^{2},  \label{8} \\
A_{3} &=&c_{3}F^{\prime }\left( c_{0}\right) +c_{1}c_{2}F^{\prime \prime
}\left( c_{0}\right) +\frac{1}{6}c_{1}^{3}F^{\prime \prime \prime }\left(
c_{0}\right)  \label{9} \\
&=&ac_{3}\left( 1-\frac{2c_{0}}{N}\right) -\frac{2a}{N}c_{1}c_{2}.
\end{eqnarray}%
Therefore, after substituting Eq. (\ref{q4}) into Eq. (\ref{q3}), the latter
becomes
\bea
\sum_{n=0}^{\infty }c_{n}\left( t,x,y,z\right) &=&c_{0}\left( x,y,z\right)
+DL_{t}^{-1}\Delta \left[ \sum_{n=0}^{\infty }c_{n}\left( t,x,y,z\right) %
\right] +\nonumber\\
&&L_{t}^{-1}\left[ \sum_{n=0}^{\infty }A_{n}\left( t,x,y,z\right) %
\right] .  \label{q10}
\eea

We rewrite Eq. (\ref{q10}) as%
\begin{equation}
c_{0}\left( t,x,y,z\right) =c_{0}\left( x,y,z\right) ,  \label{11}
\end{equation}%
\begin{equation}
c_{k+1}\left( t,x,y,z\right) =DL_{t}^{-1}\Delta \left[ c_{k}\left(
t,x,y,z\right) \right] +L_{t}^{-1}\left[ A_{k}\left( t,x,y,z\right) \right] .
\label{12}
\end{equation}%
From Eq. (\ref{12}), we obtain
\begin{equation}
c_{1}=\int_{0}^{t}\left( D\Delta c_{0}+A_{0}\right) dt,
c_{2}=\int_{0}^{t}\left( D\Delta c_{1}+A_{1}\right) dt,
c_{3}=\int_{0}^{t}\left( D\Delta c_{2}+A_{2}\right) dt,
\end{equation}

\begin{equation}
...
\end{equation}
\begin{equation}
c_{m+1}=\int_{0}^{t}\left( D\Delta c_{m}+A_{m}\right) dt.
\end{equation}%
where $k=0,1,2...m.$ The approximate solution of the Fisher-Kolmogorov
equation can be written as
\begin{equation}
c\left( t,x,y,z\right) =\sum_{i=0}^{m+1}c_{i}\left( t,x,y,z\right) .
\end{equation}

\subsection{Example: Fisher-Kolmogorov equation with the initial condition $c\left( 0,x,y,z\right) =x^{2}+y^{2}+z^{2}$}

As an example of the application of the Adomian Decomposition Method, we consider the case in which Eq. (%
\ref{q1}) should be solved together the initial condition $c\left( 0,x,y,z\right)
=c_{0}\left( x,y,z\right) =x^{2}+y^{2}+z^{2}$. Then we obtain
\begin{equation}
c_{0}\left( x,y,z\right) =x^{2}+y^{2}+z^{2},
\end{equation}
\bea
c_{1}\left( t,x,y,z\right) &=&\int_{0}^{t}\left( D\Delta c_{0}+A_{0}\right)
dt=t\Bigg[ 6D+a\left( x^{2}+y^{2}+z^{2}\right) \times \nonumber\\
&&\left( 1-\frac{%
x^{2}+y^{2}+z^{2}}{N}\right) \Bigg] ,
\eea
\begin{eqnarray}
c_{2}\left( t,x,y,z\right) &=&\int_{0}^{t}\left( D\Delta c_{1}+A_{1}\right)
dt \\
&=&\frac{at^{2}}{2N^{2}}\Bigg\{
4DN\left[ 3N-8\left( x^{2}+y^{2}+z^{2}\right) \right] +a\left(
x^{2}+y^{2}+z^{2}\right) \times \nonumber\\
&&\left[ N^{2}-3N\left( x^{2}+y^{2}+z^{2}\right) +2\left(
x^{2}+y^{2}+z^{2}\right) ^{2}\right]%
\Bigg\} ,
\end{eqnarray}
\begin{equation}
c_{3}\left( t,x,y,z\right) =\int_{0}^{t}\left( D\Delta c_{2}+A_{2}\right) dt,%
c_{4}\left( t,x,y,z\right) =\int_{0}^{t}\left( D\Delta c_{3}+A_{3}\right) dt,
\end{equation}
\begin{equation}
c\left( t,x,y,z\right) \approx c_{0}+c_{1}+c_{2}+c_{3}+c_{4}.
\end{equation}

Hence it follows that the Adomian method is also a very powerful approach for solving partial
differential equations. The solutions converge fast,  thus saving a lot of
computing time.

\section{The Laplace-Adomian Decomposition Method}\label{sect1n}

A very powerful version of the Adomian Decomposition Method is represented by the so-called Laplace-Adomian Decomposition Method (LADM) \citep{LT0, LT0a, LT1, LT2, LT3}. We will introduce this method by considering the particular example of a second order nonlinear differential equation of the form
\begin{equation}\label{eqb}
\frac{d^2y}{dx^2}+\omega ^{2}y+b^{2}+f(y)=0,
\end{equation}%
where $\omega $ and $b$ are arbitrary constants, while $f(y)$ is an nonlinear arbitrary
function of the dependent variable $y$. We will consider Eq.~(\ref{eqb}) together  with
the initial conditions $y(0)=y_{0}=a$, and $y^{\prime }(0)=0$, respectively.

We define the Laplace transform operator  $\mathcal{L}_x$ of an arbitrary function $f(x)$, as  $\mathcal{L}_x[f(x)](s) = \int_0^{\infty}{f(x)e^{-sx}dx}$.

The first, and essential step in the Laplace-Adomian Decomposition Method is to apply the Laplace transform
operator $\mathcal{L}_x$ to Eq.~(\ref{eqb}). Hence we obtain
\begin{equation}
\mathcal{L}_x\left[\frac{d^2y}{dx^2}\right] +\omega ^{2}\mathcal{L}_x[y]+%
\mathcal{L}_x[b^{2}]+\mathcal{L}_x\left[f(y)\right] =0.
\end{equation}

By using the basic properties of the Laplace transform we  straightforwardly obtain
\begin{equation}
\left( s^{2}+\omega ^{2}\right) \mathcal{L}_x[y]-sy(0)-y^{\prime }(0)+\frac{%
b^{2}}{s}+\mathcal{L}_x\left[ f(y)\right] =0.
\end{equation}

After explicitly taking into account the initial conditions for our problem we obtain the relation
\begin{equation}
\mathcal{L}_x[y]=\frac{as}{s^{2}+\omega ^{2}}-\frac{b^{2}}{s\left(
s^{2}+\omega ^{2}\right) }-\frac{1}{s^{2}+\omega ^{2}}\;\mathcal{L}_x[f(y)].
\label{6a}
\end{equation}

We assume now that the solution of Eq.~(\ref{eqb}) can be
represented in the form of an infinite series given by
\begin{equation}
y(x)=\sum_{n=0}^{\infty }y_{n}(x),  \label{7a}
\end{equation}%
where each term $y_{n}(x)$ can be calculated recursively. With respect to the nonlinear
operator $f(y)$, we assume that it can be decomposed according to
\begin{equation}
f(y)=\sum_{n=0}^{\infty }A_{n},  \label{8a}
\end{equation}%
where the functions $A_{n}$ are the Adomian polynomials, which can be obtained from the general algorithm  \citep{new0, R2}
\begin{equation}
A_{n}=\left. \frac{1}{n!}\frac{d^{n}}{d\epsilon ^{n}}f\left(
\sum_{i=0}^{\infty }{\epsilon ^{i}y_{i}}\right) \right\vert _{\epsilon =0}.
\end{equation}

The first few Adomian polynomials are given by,
\begin{equation}
A_{0}=f\left( y_0\right) ,
A_{1}=y_{1}f^{\prime }\left( y_0\right) ,
A_{2}=y_{2}f^{\prime }\left( y_0\right) +\frac{1}{2}y_{1}^{2}f^{\prime
\prime }\left( y_0\right) ,  \label{Ad2}
\end{equation}%
\begin{equation}
A_{3}=y_{3}f^{\prime }\left( y_0\right) +y_{1}y_{2}f^{\prime \prime }\left(
y_0\right) +\frac{1}{6}y_{1}^{3}f^{\prime \prime \prime }\left( y_0\right) ,
\label{Ad3}
\end{equation}
\begin{equation}
A_{4}=y_{4}f^{\prime }\left( y_0\right) +\left[ \frac{1}{2!}%
y_{2}^{2}+y_{1}y_{3}\right] f^{\prime \prime }\left( y_0\right) +\frac{1}{2!}%
y_{1}^{2}y_{2}f^{\prime \prime \prime }\left( y_0\right) +\frac{1}{4!}%
y_{1}^{4}f^{(\mathrm{iv})}\left( y_0\right) .  \label{Ad4}
\end{equation}

After substituting Eqs. (\ref{7a}) and (\ref{8a}) into Eq. (\ref{6a}) we find
\begin{equation}
\mathcal{L}_x\left[ \sum_{n=0}^{\infty }y_{n}(x)\right] =\frac{as}{%
s^{2}+\omega ^{2}}-\frac{b^2}{s\left( s^{2}+\omega ^{2}\right) }-\frac{1}{%
s^{2}+\omega ^{2}}\mathcal{L}_x[\sum_{n=0}^{\infty }A_{n}].  \label{11a}
\end{equation}

By matching both sides of Eq.~(\ref{11a}) gives an iterative
algorithm for obtaining the power series solution of Eq. (\ref{eqb}), which can be formulated as
\begin{equation}
\mathcal{L}_x\left[ y_{0}\right] =\frac{as}{s^{2}+\omega ^{2}}-\frac{b^{2}}{%
s\left( s^{2}+\omega ^{2}\right) },  \label{12a}
\end{equation}%
\begin{equation}
\mathcal{L}_x\left[ y_{1}\right] =-\frac{1}{s^{2}+\omega ^{2}}\mathcal{L}_x\left[
A_{0}\right] ,  \label{12b}
\end{equation}%
\begin{equation}
\mathcal{L}_x\left[ y_{2}\right] =-\frac{1}{s^{2}+\omega ^{2}}\mathcal{L}_x\left[
A_{1}\right] ,  \label{12c}
\end{equation}%
\begin{equation*}
...
\end{equation*}%
\begin{equation}
\mathcal{L}_x\left[ y_{k+1}\right] =-\frac{1}{s^{2}+\omega ^{2}}\mathcal{L}_x%
\left[ A_{k}\right] .  \label{12n}
\end{equation}

To obtain the value of $y_0$ we apply the inverse Laplace transform to Eq. (\ref{12a}). After substituting $y_{0}$ into the first of Eqs. (\ref{Ad2}) we find the
first Adomian polynomial $A_{0}$. The obtained expression of $A_{0}$ is then substituted into Eq. (\ref%
{12b}), which allows to compute the Laplace transforms of the quantities of its
right-hand. Then the further application of the inverse Laplace transform
gives the functional expressions of $y_{1}$. All the other terms $y_{2}$, $y_{3}$, . . ., $%
y_{k+1}, ...$ of the series solution can be similarly calculated recursively by using a step by step procedure.

\subsection{Example: $f(y)=\sum_{l=0}^{m}a_{l+2}y^{l+2}$}

We will illustrate the applications of the Laplace-Adomian Decomposition Method by considering the case of a second order nonlinear differential equation having the form \citep{PP}
\begin{equation}
\frac{d^{2}y}{dx^{2}}+\omega ^{2}y+b^{2}+\sum_{l=0}^{m}a_{l+2}y^{l+2}=0,
\label{s1}
\end{equation}
where $\omega $, $b$, and $a_{l+2}$, $l=0,...,m$ are arbitrary constants. As usual, we consider Eq.~(\ref{s1}%
) together with the set of initial conditions $y(0)=y_{0}=a$, and $%
y^{\prime }(0)=0$, respectively. We investigate Eq.~(\ref{s1}) by using the Laplace-Adomian Decomposition Method. Hence, as a first step, we
 apply the Laplace transform to Eq.~(\ref{s1}), thus finding
\begin{equation}
\mathcal{L}_x\left( \frac{d^{2}y}{dx^{2}}\right) +\omega ^{2}\mathcal{L}_x\left(
y\right) +b^{2}\mathcal{L}_x\left( 1\right) +\sum_{l=0}^{m}a_{l+2}\mathcal{L}_x%
\left[ y^{l+2}\right] =0.
\end{equation}

Next, by  the use of the properties of the Laplace transform, we immediately obtain
\begin{equation}
\mathcal{L}_x\left( y\right) \left( s^{2}+\omega ^{2}\right) =sy\left(
0\right) +y^{\prime }\left( 0\right) -\frac{b^{2}}{s}-\sum_{l=0}^{m}a_{l+2}%
\mathcal{L}_x\left[ y^{l+2}\right] =0,
\end{equation}
and thus
\begin{equation}
\mathcal{L}_x\left( y\right) =\frac{sy\left( 0\right) +y^{\prime }\left(
0\right) }{s^{2}+\omega ^{2}}-\frac{b^{2}}{s\left( s^{2}+\omega ^{2}\right) }%
-\frac{1}{s^{2}+\omega ^{2}}\sum_{l=0}^{m}a_{l+2}\mathcal{L}_x\left[ y^{l+2}%
\right] .  \label{s2}
\end{equation}

Hence, from Eq.~(\ref{s2})  $y(x)$ in obtained in the form%
\begin{equation}
y\left( x\right) =\mathcal{L}^{-1}_x\left[ \frac{sy\left( 0\right) +y^{\prime
}\left( 0\right) }{s^{2}+\omega ^{2}}-\frac{b^{2}}{s\left( s^{2}+\omega
^{2}\right) }\right] -\mathcal{L}^{-1}_x\left[ \frac{1}{s^{2}+\omega ^{2}}%
\sum_{l=0}^{m}a_{l+2}\mathcal{L}_x\left[ y^{l+2}\right] \right] .
\end{equation}

We assume now that the solution $y(x)$  of Eq.~(\ref{s1}) can be represented as $y(x)=\sum_{n=0}^{\infty
}y_{n}(x)$. Moreover, we decompose the nonlinear terms according to
\begin{equation}
y^{l+2}=\sum_{n=0}^{\infty }{A_{n,l+2}(x)},
\end{equation}%
where $A_{n,l+2}$ are the Adomian polynomials determining $y^{l+2}$.
Then we obtain
\bea \label{s3}
\sum_{n=0}^{\infty }y_{n}(x)&=&\mathcal{L}^{-1}_x\left[ \frac{sy\left( 0\right)
+y^{\prime }\left( 0\right) }{s^{2}+\omega ^{2}}-\frac{b^{2}}{s\left(
s^{2}+\omega ^{2}\right) }\right] -\nonumber\\
&&\mathcal{L}^{-1}_x\left\{ \frac{1}{%
s^{2}+\omega ^{2}}\sum_{l=0}^{m}a_{l+2}\mathcal{L}\left[ \sum_{n=0}^{\infty }%
{A_{n,l+2}(x)}\right] \right\} .
\eea

We reformulate now Eq.~(\ref{s3}) in the form
\begin{eqnarray}
y_{0}\left( x\right) +\sum_{n=0}^{\infty }y_{n+1}(x) &=&\mathcal{L}^{-1}_x%
\left[ \frac{sy\left( 0\right) +y^{\prime }\left( 0\right) }{s^{2}+\omega
^{2}}-\frac{b^{2}}{s\left( s^{2}+\omega ^{2}\right) }\right] -  \notag \\
&&\sum_{n=0}^{\infty }\mathcal{L}^{-1}_x\left\{ \frac{1}{s^{2}+\omega ^{2}}%
\sum_{l=0}^{m}a_{l+2}\mathcal{L}_x\left[ {A_{n,l+2}(x)}\right] \right\} .
\label{s4}
\end{eqnarray}

Hence from Eq.~(\ref{s4}) we obtain the set of recursive relations
\begin{equation}
y_{0}\left( x\right) =\mathcal{L}^{-1}_x\left[ \frac{sy\left( 0\right)
+y^{\prime }\left( 0\right) }{s^{2}+\omega ^{2}}-\frac{b^{2}}{s\left(
s^{2}+\omega ^{2}\right) }\right] ,
\end{equation}%
\begin{equation*}
...,
\end{equation*}

\begin{equation}
y_{k+1}(x)=-\mathcal{L}^{-1}_x\left\{ \frac{1}{s^{2}+\omega ^{2}}%
\sum_{l=0}^{m}a_{l+2}\mathcal{L}_x\left[ {A_{k,l+2}(x)}\right] \right\} .
\end{equation}

A few Adomian polynomials for the function $y^{l+2}$ are given by
\begin{equation}
A_{0,l+2}=y_{0}^{l+2},
A_{1,l+2}=(l+2)y_{1}y_{0}^{l+1},  \label{h2}
\end{equation}
\begin{equation}
A_{2,l+2}=(l+2)y_{2}y_{0}^{l+1}+(l+1)\left( l+2\right) \frac{y_{1}^{2}}{2!}%
y_{0}^{l},  \label{h3}
\end{equation}%
\begin{equation}
A_{3,l+2}=(l+2)y_{3}y_{0}^{l+1}+(l+1)\left( l+2\right)
y_{1}y_{2}y_{0}^{l}+l(l+1)\left( l+2\right) \frac{y_{1}^{3}}{3!}y_{0}^{l-1}.
\label{h4}
\end{equation}

We find the first order approximation of the solution by taking $k=0$, thus
obtaining
\begin{eqnarray}
y_{1}(x) &=&-\mathcal{L}_{x}^{-1}\left\{ \frac{\mathcal{L}_{x}\left[
\sum_{l=0}^{m}a_{l+2}A_{0,l+2}\right] }{s^{2}+\omega ^{2}}\right\} =\nonumber \\
&&
-\mathcal{L}_{x}^{-1}\left\{ \frac{\mathcal{L}_{x}\left(
a_{2}y_{0}^{2}+a_{3}y_{0}^{3}+a_{4}y_{0}^{4}+...\right) }{s^{2}+\omega ^{2}}%
\right\} .
\end{eqnarray}%
For $k=1$ we find $y_{2}(x)$ as given by%
\begin{eqnarray}
y_{2}(x) &=&-\mathcal{L}_{x}^{-1}\left\{ \frac{\mathcal{L}_{x}\left[
\sum_{l=0}^{m}a_{l+2}A_{1,l+2}\right] }{s^{2}+\omega ^{2}}\right\} =
\nonumber \\
&&-\mathcal{L}_{x}^{-1}\left\{ \frac{\mathcal{L}_{x}\left(
2a_{2}y_{1}y_{0}+3a_{3}y_{1}y_{0}^{2}+4a_{4}y_{1}y_{0}^{3}+...\right) }{%
s^{2}+\omega ^{2}}\right\} .
\end{eqnarray}%
By fixing $k$ as $k=2$ yields%
\begin{eqnarray}
\hspace{-1.1cm}&&y_{3}(x) =-\mathcal{L}_{x}^{-1}\left\{ \frac{\mathcal{L}_{x}\left[
\sum_{l=0}^{m}a_{l+2}A_{2,l+2}\right] }{s^{2}+\omega ^{2}}\right\} =-\mathcal{L}_{x}^{-1}
\nonumber \\
\hspace{-1.1cm}&&\left\{ \frac{\mathcal{L}_{x}\left[ a_{2}\left(
2y_{2}y_{0}+y_{1}^{2}\right) +3a_{3}\left(
y_{2}y_{0}^{2}+y_{1}^{2}y_{0}\right) +a_{4}\left(
4y_{2}y_{0}^{3}+6y_{1}^{2}y_{0}^{2}\right) +...\right] }{s^{2}+\omega ^{2}}%
\right\} .
\end{eqnarray}%
As a last case we take $k=3$, and thus
\begin{eqnarray}
y_{4}(x) &=&-\mathcal{L}^{-1}\Bigg\{\frac{\mathcal{L}_{x}\left[
\sum_{l=0}^{m}a_{l+2}A_{3,l+2}\right] }{s^{2}+\omega ^{2}}\Bigg\}=  \nonumber
\\
&&-\mathcal{L}^{-1}\Bigg\{\frac{1}{s^{2}+\omega ^{2}}\mathcal{L}\Bigg[%
2a_{2}\left( y_{3}y_{0}+y_{1}y_{2}\right) +a_{3}\left(
3y_{3}y_{0}^{2}+6y_{1}y_{2}y_{0}+y_{1}^{3}\right) +  \nonumber \\
&&a_{4}\left( 4y_{3}y_{0}^{3}+12y_{1}y_{2}y_{0}^{2}+4y_{1}^{3}y_{0}\right)
+...\Bigg]\Bigg\}.
\end{eqnarray}%
Hence  the truncated power series solution of Eq.~(\ref{s1}) is given by%
\begin{equation}
y\left( x\right) =\sum_{n=0}^{\infty }y_{n}(x)=y_{0}\left( x\right)
+y_{1}\left( x\right) +y_{2}\left( x\right) +y_{3}\left( x\right)
+y_{4}\left( x\right) +....
\end{equation}

\section{Astronomical and astrophysical applications }\label{sect2n}

In the present Section we consider some astronomical and astrophysical applications of the ADM. In particular, we will consider the solutions of the Kepler equation via ADM, the solutions of the Lane-Emden equation, and the study of the motion of massive particles in the Schwarzschild geometry.

\subsection{The Kepler equation}

In celestial mechanics, Kepler's equation plays an essential role in the determination of the orbit of an object evolving under the action of a central force. The Kepler equation for the hyperbolic case is \citep{Ebaid}
\be\label{Kep}
e\sin H(t)-H(t)=M(t),
\ee
where $e$ is the eccentricity of the orbit, $H(t)$ is the eccentric anomaly, $M(t) =\sqrt{\mu /a^3}(t-\tau)$ represents the mean anomaly, $\mu =GM$, while $a$ is the semi-major axis of the orbit. Moreover,  the time interval for the passage through the closest point of approach to the focus of the orbit is denoted by $\tau$. The Kepler equation (\ref{Kep}) can be transformed into the forms
\be
ey(t)-{\rm arcsinh}\; y(t)=M,
\ee
under the assumption $y(t)=\sinh H(t)$, and
\be\label{Kep1}
y(t)=\alpha +\beta {\rm arcsinh}\; y(t),
\ee
where $0\leq \alpha =M/e<\infty$, and $0\leq \beta =1/e\leq 1$, respectively. In the Adomian Decomposition Method approach to the Kepler equation one assumes that $y$ and ${\rm arcsinh}\; y$ can be decomposed as $y(t)=\sum_{n=0}^{\infty}y_n(t)$, and ${\rm arcsinh y}= \sum_{n=0}^{\infty}A_n(t)$, where $A_n(t)$ are the Adomian polynomials corresponding to ${\rm arcsinh}\; y$. After substituting the series expansions into Eq.~(\ref{Kep1}) one arrives to the following recursion relations,
\be
y_0=\alpha,
\ee
\be
y_{n+1}=\beta A_n, n=0,1,2,....
\ee

The Adomian polynomials for the function ${\rm arcsinh } y$ can be obtained as follows \citep{Ebaid},
\be
A_0 = {\rm arcsinh}\; y_0(t), A_1=\frac{y_1}{\left(1+y_0^2\right)^{1/2}}, A_2=\frac{2\left(1+y_0^2\right)y_2-y_0y_1^2}{2\left(1+y_0^2\right)^{3/2}},
\ee
\be
A_3=\frac{6\left(1+y_0^2\right)y_3-6y_0\left(1+y_0^2\right)y_1y_2+\left(2y_0^2-1\right)y_1^3}{6\left(1+y_0^2\right)^{5/2}}.
\ee

Then we obtain the following solution of the Kepler equation $\Phi (t)=\sum _{i=0}^{n-1}y_i(t)$ \citep{Ebaid},
\be
\Phi_2(t)=\alpha +\beta {\rm arcsinh}\;\alpha,
\ee
\be
\Phi_3(t)=\alpha +\beta {\rm arcsinh}\;\alpha+\frac{\beta ^2{\rm arcsinh}\;\alpha}{\left(1+\alpha ^2\right)^{12}},
\ee
\bea
\Phi _4(t)&=&\alpha +\beta {\rm arcsinh}\;\alpha+\frac{\beta ^2{\rm arcsinh}\;\alpha}{\left(1+\alpha ^2\right)^{12}}+\nonumber\\
&&\frac{2\left(1+\alpha ^2\right)^{1/2}\beta ^3{\rm arcsinh}\;\alpha-\alpha \beta ^3\left({\rm arcsinh}\;\alpha\right)^2}{2\left(1+\alpha ^2\right)^{3/2}}.
\eea

For the higher order terms in the Adomian expansion, the convergence of the series and the comparison with the full numerical solution see \citet{Ebaid}.  For the study of the elliptical Kepler problem via the ADM see \citet{Ebaid1}.

\subsection{The Lane-Emden equation}

The astrophysical properties of the static Newtonian stars can be fully characterized by the two gravitational structure equations, which are represented by the mass continuity equation, and the equation of the hydrostatic equilibrium, respectively, given by \citep{Chan, Hor,Blaga, BH}
\begin{align}
  \frac{dm(r)}{dr} &= 4 \pi \rho(r) r^{2},
  \label{5}\\
  \frac{dp(r)}{dr} &= - \frac{G m(r)}{r^{2}}\rho(r) ,
  \label{6}
\end{align}
where $\rho (r)\geq 0$ is the matter density inside the star, $p(r)\geq 0$ is the thermodynamic pressure, while $m(r)\geq 0,\forall r\geq 0$ denotes the mass inside radius $r$, respectively.
To close the system of structure equations  one should assume an equation of state for the interior stellar matter, $p=p(\rho)$, which is a functional relation between the thermodynamic pressure and the density of the matter inside the star. An important equation of state is the polytropic equation of state,
for which the pressure can be expressed as a power law of the density,
\begin{align}
  p=K\rho ^{1+1/n},
\end{align}
where $K\geq 0$ and $n$ are constants, and $n \neq 0$. After eliminating the mass function $m(r)$ between the two structure equations~(\ref{5}) and (\ref{6}), respectively, we obtain a single second order non-linear differential equation given by
\begin{align}
  \frac{1}{r^{2}}\frac{d}{dr}\left(\frac{r^{2}}{\rho }\frac{dp}{dr}\right)
  =-4\pi G\rho,
  \label{7}
\end{align}
which describes the global properties of the Newtonian star. By introducing for the density a new dimensionless variable $\theta$, so that
\begin{align}
  \rho = \rho _{c} \theta^{n},
\end{align}
where $\rho _c$ is the central density,  and $n$ is the polytropic index, we obtain for the pressure the expression $p=K\rho _{c}^{1+1/n}\theta ^{n+1}$. Next we introduce the dimensionless form of the radial coordinate $\xi $, defined as
\begin{align}
  r=\alpha \xi, \qquad
  \alpha =\sqrt{\frac{(n+1)K\rho_{c}^{1/n-1}}{4\pi G}}, \qquad
  n\neq -1.
\end{align}
In these dimensionless variables Eq.~(\ref{7}) takes the form of the Lane-Emden equation of index $n$,
\begin{align}
  \theta'' + \frac{2}{\xi} \theta' + \theta^{n} = 0.
  \label{LE}
\end{align}
 To solve the Lane-Emden equation we adopt the initial conditions $\theta(0)=1$ and $\theta'(0)=0$, respectively, where the prime represents the derivative with respect to the dimensionless independent variable $\xi $.

In the limit $n \rightarrow 0$, the Lane-Emden equation has the solution $\left.\theta(\xi)\right|_{n=0}=1-\xi^2/6$. For $n=1$, the Lane-Emden equation~(\ref{LE}) reduces to a linear ordinary differential equation,  and it has the solution $\left.\theta(\xi)\right|_{n=1}=\sin(\xi)/\xi$.  The non-linear Lane-Emden equation has only one known exact solution when $n=5$, given by $\left.\theta(\xi)\right|_{n=5}=1/\sqrt{1+\xi^2/3}$. For series solutions of the mass continuity and of the general relativistic hydrostatic equilibrium equation (the Tolman-Oppenheimer-Volkoff equation), describing the interior properties of high density compact objects, see \citet{int2} and \citet{int1}, respectively.  

\subsubsection{Solving the Lane-Emden equation via ADM}

The second order nonlinear ordinary differential equation of the form
\be
\frac{d^2y}{dx^2}+\frac{k}{x}\frac{dy}{dx}+y^m=0,
\ee
where $k>0$, is called the Lane–Emden equation of the first kind \citep{LE1}. It has to be integrated together with the initial conditions $y(0)=1$ and $y'(0)=0$, respectively. The Lane-Emden equation of the second kind is given by
\be
\frac{d^2y}{dx^2}+\frac{k}{x}\frac{dy}{dx}+e^y=0,
\ee
where $k>0$, and the equation is considered together with the initial conditions $y(0)=y'(0)=0$. However, in the following we will consider the generalized Lane-Emden equation, given by \citep{LE1}
\be\label{l1}
\frac{d^2y}{dx^2}+\frac{k}{x}\frac{dy}{dx}+f(y)=0,
\ee
where $k>0$, $f(y)$ is an arbitrary analytic function of $y$, and which should be integrated with the initial conditions $y(0)=\alpha$, and $y'(0)=0$, respectively.

\subsubsubsection{The integral formulation of the Lane-Emden equation}

 Eq.~(\ref{l1}) can be reformulated as an integral equation as follows \citep{LE1}. For $k>0$, and $k\neq 1$, Eq.~(\ref{l1}) can be reformulated as
\be
\left(x^ky'\right)'=-x^kf(y),
\ee
where a prime denotes the differentiation with respect to $x$. Integrating once we obtain
\be
y'(x)=-\frac{1}{x^k}\int_0^x{t^kf(y(t))dt}.
\ee

Integrating again we find
\be
y(x)-\alpha=-\int_0^x{\frac{1}{x^k}\int_0^x{t^kf(y(t))dtdx}}=\frac{1}{k-1}\int _0^x\int_0^x{t^kf(y(t))dtd\left(\frac{1}{x^{k-1}}\right)}.
\ee

By using the Cauchy formula for repeated integration,
\be
\int_a^x\int_a^{x_1}...\int_a^{x_{n-1}}f\left(x_n\right)dx_n...dx_2dx_1=\frac{1}{(n-1)!}\int_a^x{(x-t)^{n-1}f(t)dt},
\ee
we immediately  obtain the
integral equation formulation of the Lane-Emden equation for $k\neq 1$ as \citep{LE1}
\be
y(x)=\alpha +\frac{1}{k-1}\int_0^x{t\left(\frac{t^{k-1}}{x^{k-1}}-1\right)f(y(t))dt}, k>0, k\neq 1.
\ee

For the case $k=1$ we find \citep{LE1}
\be
y(x)=\alpha +\int _0^x{}t\ln \left(\frac{t}{x}\right)f(y(t))dt, k=1.
\ee

These two cases can be unified in a single formulation once we introduce the integral kernel $K(x,t;k)$,  defined as \citep{LE1},
\begin{equation}
K(x,t;k)=\left\{
\begin{array}{c}
\frac{1}{k-1}t\left( \frac{t^{k-1}}{x^{k-1}}-1\right) ,k>0,k\neq 1, \\
t\ln \left( \frac{t}{x}\right) ,k=1%
\end{array}%
\right. .
\end{equation}

Then the Lane-Emden equation can be formulated generally in an integral form as \citep{LE1}
\be\label{intf}
y(x)=\alpha +\int_0^x{K(x,t;k)f(y(t))dt}.
\ee

\subsubsubsection{The Adomian Decomposition Method}

As usual in the Adomian Decomposition Method, we assume that the solution $y(x)$ of the Lane-Emden equation can be represented in the form  of an infinite series, $y(x)=\sum_{n=0}^{\infty}{y_n(x)}$, while the nonlinear term $f(y)$ is decomposed by using the Adomian polynomials, $f(y(x))=\sum_{n=0}^{\infty}{A_n\left(y_0(x),y_1(x),...,y_n(x)\right)}$. Then by substituting these expressions into Eq.~(\ref{intf}) we find
\be
\sum_{n=0}^{\infty}{y_n(x)}=\alpha +\int_0^x{K(x,t;k)\sum_{n=0}^{\infty}{A_n\left(y_0(t),y_1(t),...,y_n(t)\right)}dt}, k>0.
\ee

By choosing $y_0=\alpha$, we find the following set of recursive relations for the terms in the series solution of the Lane-Emden equation,
\be
y_0=\alpha,
\ee
\be
y_{m+1}=\int_0^x{K(x,t;k)A_m\left(y_0(t),y_1(t),...,y_n(t)\right)dt},m\geq 0.
\ee

The above set of relations will lead to the complete determination of each of the components $y_n(x)$ of the solution $y(x)$. As a simple application of the Adomian Decomposition Method to the nonlinear Lane-Emden type equations we consider, following \citet{LE1}, the case of the equation
\be
\frac{d^2y}{dx^2}+\frac{1}{x}\frac{dy}{dx}-16x^2e^{-2y}=0,y(0)=0,y'(0)=0,
\ee
which has the exact solution
\be\label{exact1}
y(x)=\ln \left(1+x^4\right).
\ee
The recursive Adomian relation is obtained as
\be
y_0(x)=0,
\ee
\be
y_{m+1}(x)=\int_0^x{t\ln\left(\frac{t}{x}\right)\left(-16t^2A_m(t)\right)dt}, m\geq 0.
\ee
After computing the Adomian polynomials for the nonlinear term $e^{-2y}$, we obtain
\be
y_0(x)=0, y_1(x)=x^4, y_2(x)=-\frac{1}{2}x^8, y_3(x)=\frac{1}{3}x^{12}, y_4(x)=-\frac{1}{4}x^{16}, ....
\ee

It is easy to see by series expanding the exact solution (\ref{exact1}) that the Adomian series solution
\be
y(x) = x^4 - \frac{1}{2}x^8 + \frac{1}{3}x^{12} - \frac{1}{4}x^{16} +...,
\ee
coincides with the exact solution.

\subsection{Solving the equation of motion of celestial objects
in Schwarzschild geometry}

  The equation of motion describing the general relativistic motion of a massive celestial body (for example, a planet) in the spherically symmetric and static Schwarzschild geometry, written in spherical coordinates $\left(r,\varphi,\theta\right)$, is given by
\begin{equation}
\frac{d^{2}u}{d\varphi ^{2}}+u=\frac{M}{L^{2}}+3Mu^{2},  \label{a}
\end{equation}%
where $u=1/r$. For the details of Schwarzschild geometry and of the derivation of Eq.~(\ref{a}) see \citet{PP} and \citet{HL}. In the following we use the natural system of units with $G=c=1$.

To obtain a simpler mathematical formalism we rescale the function $u=1/r$ according to
\begin{equation}
u=\frac{1}{3M}U.
\end{equation}%
Thus Eq.~(\ref{a}) takes the form
\begin{equation}  \label{a1}
\frac{d^{2}U}{d\varphi ^{2}}+U=b^{2}+U^{2},
\end{equation}%
where we have denoted $b^{2}=3M^{2}/L^{2}$. We will consider Eq.~(\ref{a1}) together
with the initial conditions $U(0)=3Mu(0)=a$, and $U^{\prime }(0)=0$,
respectively. In the following we will obtain semi-analytical solutions of Eq.~(\ref{a1}) by using the Laplace-Adomian Method.

\subsubsection{Power series solution of the equation of motion}

 We assume that the solution of Eq. (\ref{a1}) can be obtained in the form of a power series, so that
\begin{equation}  \label{62}
U\left( \varphi \right) =\sum_{n=0}^{\infty }U_{n}\left( \varphi \right) .
\end{equation}

We apply now the Laplace transform operator $\mathcal{L}_{\varphi}$ to Eq. (\ref{a1}), thus obtaining
\begin{equation}
\mathcal{L}_{\varphi}\left[ \frac{d^{2}U}{d\varphi ^{2}}\right] +\mathcal{L}_{\varphi}\left[ U%
\right] =b^2\mathcal{L}_{\varphi}\left[ 1\right] +\mathcal{L}_{\varphi}\left[ U^{2}\right] .
\end{equation}

By using the properties of the Laplace transform we find
\begin{equation}
s^{2}\mathcal{L}_{\varphi}\left( U\right) -sU\left( 0\right) -U^{\prime }\left(
0\right) +\mathcal{L}_{\varphi}\left( U\right) =\frac{b^{2}}{s}+\mathcal{L}_{\varphi}\left[ U^{2}%
\right] ,
\end{equation}
and
\begin{equation}
\mathcal{L}_{\varphi}\left( U\right) =\frac{sU\left( 0\right) +U^{\prime }\left(
0\right) }{s^{2}+1}+\frac{b^{2}}{s\left( s^{2}+1\right) }+\frac{1}{s^{2}+1}%
\mathcal{L}_{\varphi}\left[ U^{2}\right],   \label{a2l}
\end{equation}
respectively.
The first four Adomian polynomials for $U^{2}$ are given by
\begin{equation}
A_{0}=U_{0}^{2},
A_{1}=2U_{1}U_{0},
A_{2}=2U_{2}U_{0}+U_{1}^{2},
A_{3}=2U_{3}U_{0}+2U_{1}U_{2}.
\end{equation}%
Now we substitute Eq. (\ref{62}) and $U^{2}=\sum_{n=0}^{\infty }A_{n}\left(
\varphi \right) $ into Eq. (\ref{a2l}), and thus we obtain the relation
\begin{equation}
\mathcal{L}_{\varphi}\left[ \sum_{n=0}^{\infty }U_{n}\left( \varphi \right) \right] =%
\frac{sU\left( 0\right) +U^{\prime }\left( 0\right) }{s^{2}+1}+\frac{b^{2}}{%
s\left( s^{2}+1\right) }+\frac{1}{s^{2}+1}\mathcal{L}_{\varphi}\left[
\sum_{n=0}^{\infty }A_{n}\left( \varphi \right) \right] ,
\end{equation}%
which can be written explicitly as
\begin{eqnarray}  \label{a2}
&&U_{0}\left( \varphi \right) +\sum_{n=1}^{\infty }U_{n}\left( \varphi \right)
=U_{0}\left( \varphi \right) +\sum_{n=0}^{\infty }U_{n+1}\left( \varphi
\right) =\nonumber\\
&&\mathcal{L}^{-1}_{\varphi}\left[ \frac{sU\left( 0\right) +U^{\prime }\left(
0\right) }{s^{2}+1}+\frac{b^{2}}{s\left( s^{2}+1\right) }\right] + \sum_{n=0}^{\infty }\mathcal{L}^{-1}_{\varphi}\left[ \frac{\mathcal{L}_{\varphi}\left[
A_{n}\left( \varphi \right) \right] }{s^{2}+1}\right] .
\end{eqnarray}

Now we can rewrite Eq. (\ref{a2}) in the following recursive forms
\begin{equation}
U_{0}\left( \varphi \right) =\mathcal{L}^{-1}_{\varphi}\left[ \frac{sU\left( 0\right)
+U^{\prime }\left( 0\right) }{s^{2}+1}+\frac{b^{2}}{s\left( s^{2}+1\right) }%
\right] ,
\end{equation}%
\begin{equation*}
...,
\end{equation*}
\begin{equation}
U_{k+1}\left( \varphi \right) =\mathcal{L}^{-1}_{\varphi}\left[ \frac{\mathcal{L}_{\varphi}\left[
A_{k}\left( \varphi \right) \right] }{s^{2}+1}\right] .
\end{equation}

\subsubsubsection{The explicit terms of the Adomian expansion}

 By using the explicit expressions of the Adomian polynomials, we can find the analytical forms of the successive terms in the Adomian series expansion of the solution of the general relativistic equation of motion of a planet in the spherically symmetric and static Schwarzschild geometry as follows.  First of all, by neglecting the nonlinear term in Eq.~(\ref{a1}), we obtain the zeroth order approximation of the solution as given by
\begin{equation}
U_{0}(\varphi )=\left( a-b^{2}\right) \cos \varphi +b^{2}.
\end{equation}%
Then for the first Adomian polynomial we obtain
\begin{equation}
A_{0}=U_{0}^{2}=\left[ \left( a-b^{2}\right) \cos \varphi +b^{2}\right] ^{2},
\end{equation}%
Once $A_0$ is known, for the first order approximation of the solution we find
\begin{equation}
U_{1}\left( \varphi \right) =\mathcal{L}^{-1}_{\varphi}\left[ \frac{\mathcal{L}_{\varphi}\left[
A_{0}\left( \varphi \right) \right] }{s^{2}+1}\right] =\mathcal{L}^{-1}_{\varphi}\left[
\frac{\mathcal{L}_{\varphi}\left[ U_{0}^{2}\right] }{s^{2}+1}\right] ,
\end{equation}%
or, explicitly,
\begin{eqnarray}
U_{1}(\varphi ) &=&\frac{1}{6}\Bigg\{-2\left( a^{2}-2ab^{2}+4b^{4}\right)
\cos \varphi +3\left( a^{2}-2ab^{2}+3b^{4}\right) +  \notag \\
&&\left( a-b^{2}\right) \left[ \left( b^{2}-a\right) \cos (2\varphi
)+6b^{2}\varphi \sin \varphi \right] \Bigg\},
\end{eqnarray}
The Adomian polynomial $A_1$ can be then obtained as
\begin{eqnarray}
\hspace{-0.8cm}A_{1} &=&2U_{1}U_{0}=\frac{1}{3}\Bigg[\left( a-b^{2}\right) \cos \varphi
+b^{2}\Bigg]\Bigg\{-2\left( a^{2}-2ab^{2}+4b^{4}\right) \cos \varphi +
\notag \\
\hspace{-0.8cm}&&3\left( a^{2}-2ab^{2}+3b^{4}\right) +\left( a-b^{2}\right) \left[ \left(
b^{2}-a\right) \cos (2\varphi )+6b^{2}\varphi \sin \varphi \right] \Bigg\},
\end{eqnarray}
giving for the second order approximation the expression
\begin{equation}
U_{2}\left( \varphi \right) =\mathcal{L}^{-1}_{\varphi}\left[ \frac{\mathcal{L}_{\varphi}\left[
A_{1}\left( \varphi \right) \right] }{s^{2}+1}\right] =2\mathcal{L}^{-1}_{\varphi}%
\left[ \frac{\mathcal{L}_{\varphi}\left[ U_{1}U_{0}\right] }{s^{2}+1}\right] ,
\end{equation}
or, explicitly,
\begin{eqnarray}
\hspace{-0.5cm}&&U_{2}\left( \varphi \right) =\frac{1}{144}\Bigg\{16\left(
a^{2}-5ab^{2}+7b^{4}\right) \left( a-b^{2}\right) \cos (2\varphi )+\cos
\varphi \Bigg[29a^{3}-183a^{2}b^{2}+  \notag \\
\hspace{-0.5cm}&&3ab^{4}\left( 125-24\varphi ^{2}\right) +b^{6}\left( 72\varphi
^{2}-509\right) \Bigg]+12\varphi \left(
5a^{3}-19a^{2}b^{2}+41ab^{4}-39b^{6}\right) \times  \notag \\
\hspace{-0.5cm}&&\sin \varphi -48\left( a^{3}-6a^{2}b^{2}+12ab^{4}-13b^{6}\right)
-48\varphi \left( b^{3}-ab\right) ^{2}\sin (2\varphi )+  \notag \\
\hspace{-0.5cm}&&3\left( a-b^{2}\right) ^{3}\cos (3\varphi )\Bigg\},
\end{eqnarray}%
For the higher terms expansions of the solutions of the general relativistic equation of motion of a massive celestial object in Schwarzschild geometry see \citet{PP}, where astrophysical applications of the method (motion of the planet Mercury, perihelion precession, and light deflection) are also presented, and discussed in detail. The study of the deflection of light can be done in a similar manner. Generally, by using LADM  we can obtain  the power series representation of the solution of the general relativistic equation of motion of planets in Schwarzschild geometry up to an arbitrary precision level as
$U\left( \varphi \right) =U_{0}\left( \varphi \right) +U_{1}\left( \varphi
\right) +U_{2}\left( \varphi \right) +U_{3}\left( \varphi \right)
+U_{4}\left( \varphi \right) +....$.

\section{Discussions and concluding remarks}\label{sect6}

In the present paper we have presented, at an introductory level, some aspects of the powerful method introduced by G. Adomian to solve nonlinear differential, stochastic and functional equations. Usually this method is known as the Adomian Decomposition Method, or ADM for short. The mathematical technique is essentially based on the decomposition of the solution of the nonlinear operator equation into a series of analytic functions. Each term of the Adomian decomposition series is computed from a polynomial obtained from the power series expansion of an analytic function. The Adomian technique is very simple, efficient, and effective, but, on the other hand, it may raise the necessity of the in depth investigations of the convergence of the series of functions representing the solution of the given nonlinear equation \cite{C1,C2}. The Adomian Decomposition Method has been used very successfully to obtain semianalytical solutions  for many important classes of functional, differential, and integral  equations, respectively, with important applications in many fields of fundamental and applied sciences, and engineering, respectively. The key to the success of the method relies in the decomposition of the nonlinear term in the differential or integral equations into a series of polynomials of the form $\sum _{n=1}^{\infty}{A_n}$, where $A_n$ are polynomials known as the  Adomian polynomials. A large number of algorithms and formulas that can calculate the Adomian polynomials for all expressions of nonlinearity were introduced in \cite{R1,R2}.

Even that the Adomian method is discussed in many articles, a systematic, simple and pedagogical introduction to the subject is still missing. It is the main goal of the present paper to provide such an introduction, which may be useful for scientists who would like to learn about this method by investigating its simplest applications, before proceeding to more advanced topics. After introducing the basics of the method, we have discussed in detail the ADM for the standard differential equations of mathematics, including the linear ordinary differential equation, and the Bernoulli, Riccati and Abel equations, respectively. In each case we have described in detail the general formalism and the particular method, and we have written down explicitly the Adomian form of the solution. For each type of considered equations we have also analyzed a concrete example, and we have shown that the Adomian solution exactly coincides with the analytic solution that can be obtained by using standard mathematical methods. This full agreement explicitly indicates the power of the Adomian Decomposition Method, which could lead to obtaining even the exact solution of a given complicated nonlinear ordinary differential equation, or of an integral equation. We have performed a similar analysis for the second order and the fifth order ordinary differential equations, by explicitly formulating the full process of obtaining the series solution. Specific example have been analyzed for each case. A very powerful extension of the ADM, the Laplace-Adomian Decomposition Method was also introduced through the study of a particular example of a second order nonlinear differential equation.

Finally, we have briefly considered the applications of the Adomian Decomposition Method to some important cases of differential equations that play an essential role in physics and astronomy. Thus, we have presented in detail the important case of the Fisher-Kolmogorov equation, a fundamental equation in several fields of biology, medicine and population dynamics. In this case, after presenting the general algorithm for the solution, a particular example has been investigated in detail. We have also described the applications of the ADM in three important fields of astronomy and astrophysics, namely, the determination of the orbits of celestial objects from the Kepler equation, obtaining the solutions of the nonlinear Lane-Emden equation, which plays a fundamental role in the study of the stellar structure, and for the investigation of the general relativistic motion of celestial objects in the Schwarzschild geometry. In all these fields the Adomian Decomposition Method has proven to be a computationally efficient and a highly precise theoretical tool for solving the complicated nonlinear equations describing astronomical and astrophysical phenomena.

Certainly the Adomian decomposition method represents a valuable tool for physicists and engineers working with
real physical problems. Hopefully the present introduction to this subject will determine scientists working in various fields to become more involved in this interesting and fertile field of investigation, which is very efficient and productive in dealing with large classes of differential/integral equations and complicated mathematical models describing natural phenomena.


\makeatletter
\def\@biblabel#1{}
\makeatother

\received{\it *}
\end{document}